\documentclass[10pt,twoside,twocolumn,english,aps,prl,superscriptaddress,notitlepage]{revtex4-2}
\usepackage[utf8]{inputenc}
\usepackage[a4paper]{geometry}
\usepackage{color,xcolor,soul}
\usepackage{verbatim,textcomp,enumitem}
\usepackage{amstext,amsfonts,amsmath,amssymb}
\usepackage{graphicx}
\usepackage[calcwidth,explicit]{titlesec}
\usepackage{chngcntr}
\usepackage[normalem]{ulem}
\usepackage{soul}
\usepackage[textsize=scriptsize,textwidth=2cm]{todonotes}
\usepackage[unicode=true,pdfusetitle, bookmarks=false, breaklinks=true,pdfborder={0 0 0},pdfborderstyle={},backref=false,colorlinks=true]{hyperref}
\hypersetup{colorlinks=true,citecolor=blue,linkcolor=blue,urlcolor=blue}

\makeatletter\renewcommand\frontmatter@abstractwidth{\dimexpr\textwidth-2cm\relax}\makeatother

\titleformat{\section}{}{}{0pt}{}
\titleformat{\section}{\bfseries\sffamily\filcenter}{}{0.2em}{#1}
\titlespacing{\section}{0pt}{0.2ex}{0.1ex}
\titleformat{\paragraph}[runin]{\normalfont\normalsize\bfseries}{}{0pt}{}
\titlespacing*{\paragraph}{0em}{0ex}{0.5em}[]

{}

\setcitestyle{super}

\counterwithout{paragraph}{subsubsection}

\AtBeginDocument{
	\renewcommand{\ref}[1]{\autoref{#1}}

}

\begin{document}
\title{All-Electrical Skyrmionic Bits in a Chiral Magnetic Tunnel Junction 
\smallskip{}}
\author{Shaohai Chen}\thanks{These authors contributed equally}
\affiliation{Institute of Materials Research \& Engineering, Agency for Science, Technology \& Research, 138634 Singapore}
\author{Pin Ho}\thanks{These authors contributed equally}
\affiliation{Institute of Materials Research \& Engineering, Agency for Science, Technology \& Research, 138634 Singapore}
\author{James Lourembam}\thanks{These authors contributed equally}
\affiliation{Institute of Materials Research \& Engineering, Agency for Science, Technology \& Research, 138634 Singapore}
\author{Alexander K.J. Toh}
\affiliation{Institute of Materials Research \& Engineering, Agency for Science, Technology \& Research, 138634 Singapore}
\author{Jifei Huang}
\affiliation{Physics Department, National University of Singapore, 117551 Singapore}
\author{Xiaoye Chen}
\affiliation{Institute of Materials Research \& Engineering, Agency for Science, Technology \& Research, 138634 Singapore}
\author{Hang Khume Tan}
\affiliation{Institute of Materials Research \& Engineering, Agency for Science, Technology \& Research, 138634 Singapore}
\author{Sherry K.L. Yap}
\affiliation{Institute of Materials Research \& Engineering, Agency for Science, Technology \& Research, 138634 Singapore}
\author{Royston J.J. Lim}
\affiliation{Institute of Materials Research \& Engineering, Agency for Science, Technology \& Research, 138634 Singapore}
\author{Hui Ru Tan}
\affiliation{Institute of Materials Research \& Engineering, Agency for Science, Technology \& Research, 138634 Singapore}
\author{T.S. Suraj}
\affiliation{Physics Department, National University of Singapore, 117551 Singapore}
\author{Yeow Teck Toh}
\affiliation{Institute of Materials Research \& Engineering, Agency for Science, Technology \& Research, 138634 Singapore}
\author{Idayu Lim}
\affiliation{Institute of Materials Research \& Engineering, Agency for Science, Technology \& Research, 138634 Singapore}
\author{Jing Zhou}
\affiliation{Institute of Materials Research \& Engineering, Agency for Science, Technology \& Research, 138634 Singapore}
\author{Hong Jing Chung}
\affiliation{Institute of Materials Research \& Engineering, Agency for Science, Technology \& Research, 138634 Singapore}
\author{Sze Ter Lim}
\affiliation{Institute of Materials Research \& Engineering, Agency for Science, Technology \& Research, 138634 Singapore}
\author{Anjan Soumyanarayanan}\thanks{anjan@imre.a-star.edu.sg}
\affiliation{Institute of Materials Research \& Engineering, Agency for Science, Technology \& Research, 138634 Singapore}
\affiliation{Physics Department, National University of Singapore, 117551 Singapore}

\begin{abstract}
Topological spin textures such as magnetic skyrmions hold considerable promise as robust, nanometre-scale, mobile bits for sustainable computing.
A longstanding roadblock to unleashing their potential is the absence of a device enabling deterministic electrical readout of individual spin textures.
Here we present the wafer-scale realization of a nanoscale chiral magnetic tunnel junction (MTJ) hosting a single, ambient skyrmion.
Using a suite of electrical and multi-modal imaging techniques, we show that the MTJ nucleates skyrmions of fixed polarity, whose large readout signal -- 20-70\% relative to uniform states -- corresponds directly to skyrmion size. 
Further, the MTJ exploits complementary mechanisms to stabilize distinctly sized skyrmions at zero field, thereby realizing three nonvolatile electrical states. 
Crucially, it can write and delete skyrmions using current densities 1,000 times lower than state-of-the-art.
These results provide a platform to incorporate readout and manipulation of skyrmionic bits across myriad device architectures, and a springboard to harness chiral spin textures for multi-bit memory and unconventional computing.
\end{abstract}
\maketitle

\section{Introduction}\label{sec:Intro}

\paragraph{Intro: MTJs}
Magnetic tunnel junctions (MTJs) are two-terminal devices whose resistive state is described by the orientation of an active free layer (FL).
Preferential tunnelling of spin polarized electrons 
between the FL and a fixed reference layer (RL) -- known as tunnel magnetoresistance (TMR) -- enables the detection of relative FL orientation \cite{Julliere.1975, Moodera.1995}, while current-induced spin torques are employed for FL manipulation \cite{Slonczewski.1996}.
Seminal advances, including the use of MgO as the tunnel barrier \cite{Yuasa.2004, Parkin.2004}, and CoFeB as the FL and RL \cite{Ikeda.2010}, led to nanoscale MTJs with large ambient TMR ratio, perpendicular easy axis, and wafer-scale manufacturability, paving the way for commercial magnetic random-access memory \cite{Engel.2005, Chappert.2007}. 
Recent efforts to employ MTJs for sustainable computing have prompted explorations of device characteristics inherently suited to such applications \cite{Dieny.2020}.

\paragraph{Intro: Skyrmions}
Magnetic skyrmions – nanoscale spin textures formed in industry-friendly chiral multilayer films \cite{MoreauLuchaire.2016,Woo.2016, Boulle.2016, Soumyanarayanan.2017} – have emerged as a promising avenue to realize scalable spintronic elements. 
Their topological structure gives rise to distinct dynamic properties \cite{Nagaosa.2013} and efficient electrical generation and manipulation in devices \cite{Jiang.2015, Woo.2016}.
These unique attributes have considerable promise towards non-Boolean logic and unconventional computing \cite{Zazvorka.2019, Grollier.2020}. 
However, the absence of deterministic, electrical detection of an individual skyrmion has been a longstanding impediment to its practical relevance \cite{Fert.2017}. 

\paragraph{Skyrmion TMR: Lit Review} 
Early cryogenic demonstration of magnetoresistance (MR) detection and spin current switching of skyrmions by scanning tunnelling microscopy\cite{Romming.2013} generated considerable promise towards ambient all-electrical devices. 
Within an MTJ, a skyrmion of diameter $d_{\rm S}$ should present a distinct conductance channel with magnitude proportional to its extent, $\left. d^2_{\rm S} \right/ W^2_{\rm Cell}$, where $W_{\rm Cell}$ is the MTJ cell diameter. 
Numerous skyrmionic device proposals aim to harness MTJs over a range of architectures to realize myriad computing applications \cite{Roy.2019}. 
However, experimental efforts using chiral multilayers have been restricted to micro-scale MTJs \cite{Guang.2022, Li.2022} hosting multi-domain states, wherein individual skyrmions cannot contribute sufficient or distinct MR for reliable detection. 
Conversely, efforts on nanoscale MTJs utilized achiral or weakly chiral FLs \cite{Kasai.2019, Penthorn.2019}, wherein skyrmions are not expected to be thermodynamically stable \cite{Nagaosa.2013, Fert.2017}. 
Engineering a chiral FL to host skyrmions, while also optimising MR for MTJ readout presents considerable challenges. 
Deterministic individual detection further requires skyrmion stability within MTJs of comparable size, with concomitant electrical detection and magnetic imaging.

\paragraph{Results Summary}
Here, we present a nanoscale, chiral MTJ hosting a single Néel skyrmion in addition to uniformly magnetized (UM) states.
Quantitative electrical and imaging studies reveal skyrmionic MR contribution of 20-70\%, which vary systematically with skyrmion size. 
The MTJ hosts skyrmions of one fixed polarity, with two distinct nucleation and annihilation mechanisms, thereby stabilizing multiple skyrmion states at zero field (ZF). 
The device can write and delete skyrmions with unprecedented efficiency, unleashing their potential for unconventional computing. 

\section{Stack Design and Characterization}\label{sec:StackDesign}

\begin{figure*}
    \centering \includegraphics[width=0.84\linewidth]{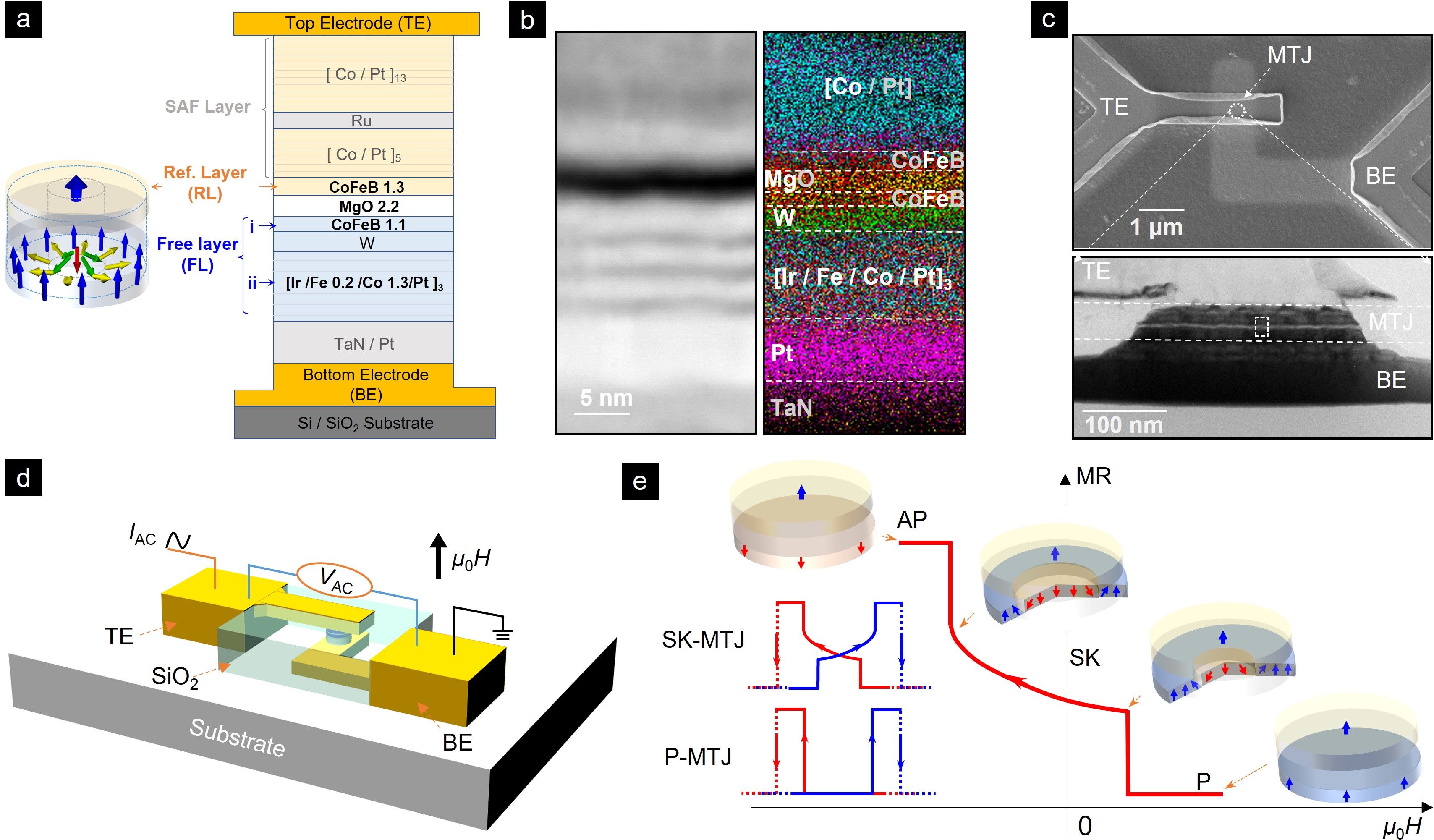} 
    \caption[Device Structure and Electrical Setup]
    {\textbf{Device Structure and Electrical Setup.} 
    \textbf{(a)} Multilayer stack structure for skyrmionic (SK)-MTJ (thicknesses in nm, see Methods). The composite free layer (FL) consists of CoFeB (i) coupled to [Ir/Fe/Co/Pt]$_3$ (ii). Other active layers are similar to a perpendicular (P)-MTJ. 
    Left inset: schematic of a Néel skyrmion in the FL of an SK-MTJ. 
    \textbf{(b-c)} Material characterization of a representative $W_{\rm Cell} \simeq 300$\,nm SK-MTJ device. 
    (b) Zoomed-in cross-sectional TEM images of active device stack (c.f. a), acquired in dark field (left) and elemental mapping (right) modalities. 
    (c) Plan view SEM image (top) of device, and cross-sectional TEM image (bottom) of the full stack. Dashed lines identify zoom-in portion for (b). 
    \textbf{(d)} Schematic of setup used for electrical measurements of MTJs, with varying OP magnetic fields, $\mu_0 H$. 
    \textbf{(e)} Schematic MR evolution with field expected for SK-MTJ device with up-oriented ($\uparrow$) RL (\ref{Eq:SKMTJ-NormMR}). In addition to uniform UM$\uparrow$ (parallel, P) and UM$\downarrow$ (anti-parallel, AP) states, we expect an intermediate skyrmion (SK) state. 
    Inset compares expected major loops for SK-MTJ and conventional P-MTJ.}
    \label{fig:Device}
\end{figure*}

\paragraph{Stack Design}
\ref{fig:Device}a shows the stack structure for the skyrmionic (SK-) MTJ, which utilizes CoFeB for the two active ferromagnetic layers sandwiching the MgO tunnel barrier (see Methods). 
The W/CoFeB/MgO stack displays perpendicular magnetic anisotropy (PMA) and sizable TMR, $R_{\rm AP}/R_{\rm P} - 1 \sim 0.95$ \cite{Kim.2017}, where $R_{\rm AP}\,(R_{\rm P})$ are the resistances of the anti-parallel, AP (parallel, P) MTJ states  (SM §S1,2). 
Nanoscale spin textures are stabilized in the bottom CoFeB via an [Ir/Fe/Co/Pt]$_3$ chiral underlayer, known to host strong interfacial Dyzaloshinskii-Moriya interaction ($1.3$~mJ/m$^2$, Methods) and robust skyrmions \cite{Soumyanarayanan.2017, Chen.2022}. 
These two layers (\ref{fig:Device}a: i, ii) are anisotropy-matched and conjoined by strong interlayer exchange coupling ($0.93$\,mJ/m$^2$, Methods) to ensure their behaviour as one \emph{\textbf{composite free layer}} (FL). 
Lorentz transmission electron microscopy (LTEM) of the FL shows a labyrinthine ZF state, with $\sim 100$~nm Néel skyrmions at polarized and unpolarized field ends (SM §S4). 
Meanwhile, the RL is coupled to a hard PMA Co/Pt pin layer, which maintains RL uniformity over the field range of interest \cite{Ikeda.2010, Han.2015}. 
The compensating synthetic antiferromagnet (SAF) overlayer ensures negligible influence of RL on FL characteristics\cite{Lim.2015}, verified by the LTEM-imaged field evolution of spin textures (SM §S3,4). 
The MTJ stack is sandwiched by top and bottom electrodes for electrical characterization.

\paragraph{Device Fabrication \& Characterization}
Two-terminal MTJ devices were fabricated using a 200~mm fabrication line (see Methods), resulting in $>90$\% yield of devices with $W_{\rm Cell}$ over 200–1000~nm.
In addition to SK-MTJs, conventional P-MTJs were also concomitantly developed for comparative purposes using a similar, yet stronger PMA stack. 
A suite of electron and force microscopy and spectroscopy tools were used for compositional, structural, and device size verifications. 
Layer-resolved TEM imaging (\ref{fig:Device}b) of a representative $W_{\rm Cell} \simeq 300$~nm MTJ device reveals discrete FL and RL sandwiching a uniform MgO barrier, with moderate taper across the active stack (\ref{fig:Device}c: bottom). 
Arrays of dot nanostructures (diameter: $W_{\rm Dot}$) were fabricated on companion SK-MTJ and FL wafers (SM §S1) for magnetic force microscopy (MFM) imaging, enabling direct comparison with electrical measurements.

\paragraph{MTJ Electrical Measurements}
Electrical resistance ($R$) measurements were performed using standard AC lock-in technique by recording the MTJ voltage in response to applied $\sim$nA currents across varying out-of-plane (OP) magnetic fields (\ref{fig:Device}d, Methods).
In addition to the P and AP states for conventional P-MTJs, the SK-MTJ should host a skyrmionic (SK) state. 
For RL oriented up (\ref{fig:Device}e), reducing $H$ from the P-state would nucleate an oppositely (down) polarized skyrmion \cite{Ho.2019}. 
The normalized MR of the SK-MTJ can be expressed using the parallel resistor model \cite{Chen.2022b} (see Methods) as
\begin{equation}\label{Eq:SKMTJ-NormMR} 
{\rm MR}(H) =\left[\frac{R_{\rm AP}}{R_{\rm P}}\, 
\left(\frac{W^2_{\rm Cell}}{d^2_{\rm S}(H)} -1 \right) + 1 \right]^{-1}\quad,
\end{equation}
Further $H$ reduction would increase the skyrmion size, and its MR, eventually saturating at the AP state.
In principle, opposite swept field should result in the converse evolution, producing a symmetric ${\rm MR}(H)$ loop.
\begin{figure*}
    \centering \includegraphics[width=1\linewidth]{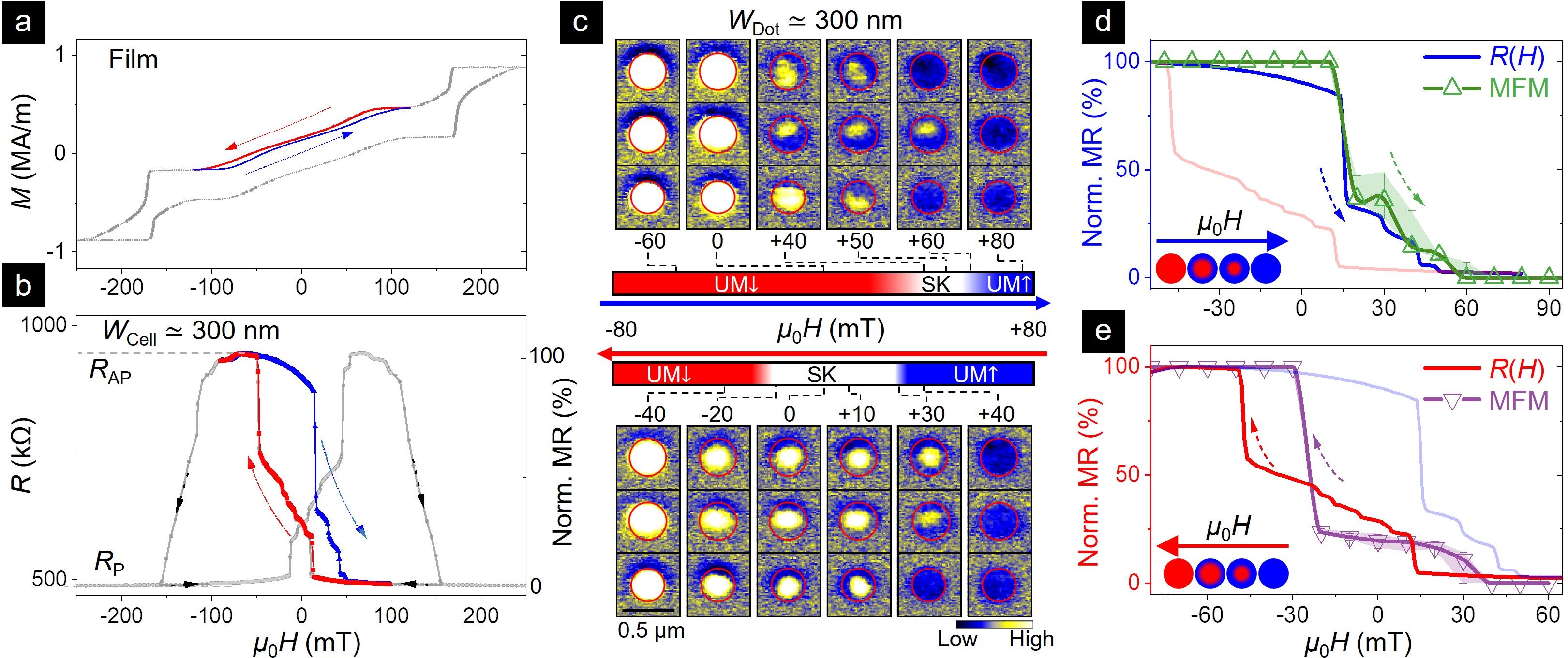} 
    \caption[Field Evolution of MR and Imaged States]
    {\textbf{Field Evolution of MR and Imaged States.} 
    \textbf{(a)} OP magnetization hysteresis loops, $M(H)$ for SK-MTJ thin film, showing major (grey) and minor (blue: up-sweep, red: down-sweep) loops. 
    \textbf{(b)} $R(H)$ loops for SK-MTJ device ($W_{\rm Cell} \simeq 300$~nm) in measured (left) and normalized MR (right, \ref{Eq:Norm-MR}) units, showing major (grey) and minor (blue: up-sweep, red: down-sweep) loops. Arrows indicate sweep directions for (a-b).  
    \textbf{(c)} MFM-imaged evolution of 3 SK-MTJ dots ($W_{\rm Dot} \simeq 300$~nm, data referenced to positive tip magnetization, see Methods) with varying \emph{in situ} $\mu_0 H$ for up-sweep (top) and down-sweep (bottom). Field recipes follow (b); red circles are guides-to-the-eye for dot positions. Centre: colour scalebar indicates statistically determined magnetization states (UM$\downarrow$: red; UM$\uparrow$: blue; SK: white). 
    \textbf{(d-e)} Comparison of normalized MR of the device (b), with the estimate from MFM-imaged median magnetic states (c: using \ref{Eq:SKMTJ-NormMR}, Methods), for (d) up-sweep and (e) down-sweep. Faint lines show opposite $R(H)$ sweep, cartoon insets show FL evolution.}
    \label{fig:TMR-Expts}
\end{figure*}

\section{TMR of Chiral MTJ\label{sec:TMR-Expts}}

\paragraph{TMR: RH loop}
We begin by examining the major (full stack) and minor (FL) hysteresis loops for the SK-MTJ stack and device. 
The film-level magnetization ($M(H)$, \ref{fig:TMR-Expts}a) shows distinct, symmetric steps due to the switching of the FL ($\sim 80$ mT) and RL ($\sim 150$ mT) respectively. 
Similarly, the major $R(H)$ loop of the prototypical SK-MTJ device ($W_{\rm Cell} \simeq 300$ nm, \ref{fig:TMR-Expts}b: grey) shows distinct P and AP states, with sizable TMR ratio ($\sim 0.95$). 
The excellent correspondence of P and AP states between the major (grey) and minor (red, blue) loops across \ref{fig:TMR-Expts}a-b confirms the limited influence of FL on the RL, and RL uniformity over the field range of interest ($\pm 100$ mT). 
Meanwhile, in contrast to the sharp switching for P-MTJs (SM §S8), the minor $R(H)$ loop for the SK-MTJ has a distinctive intermediate resistance state (\ref{fig:TMR-Expts}b), consistent with the expected formation of FL skyrmions (c.f. \ref{fig:Device}e).  
Moreover, the $R(H)$ (device) minor loop exhibits three distinguishing features compared to the $M(H)$ (film) loop: (1) reduced switching field 
and increased coercivity, 
due to demagnetization effects \cite{Ho.2019}, (2) sharp transitions to/from P, AP states, due to discrete texture nucleation, 
and (3) large $y$-axis asymmetry, i.e., larger jumps to/from AP c.f. P state. 
Crucially, such $y$-axis asymmetry for the device is at sharp variance with the symmetric $M(H)$ loop of the parent film, and is unreported in MTJ literature \cite{Ikeda.2010, Guang.2022, Li.2022}. 

\begin{figure}
	\centering \includegraphics[width=1\linewidth]{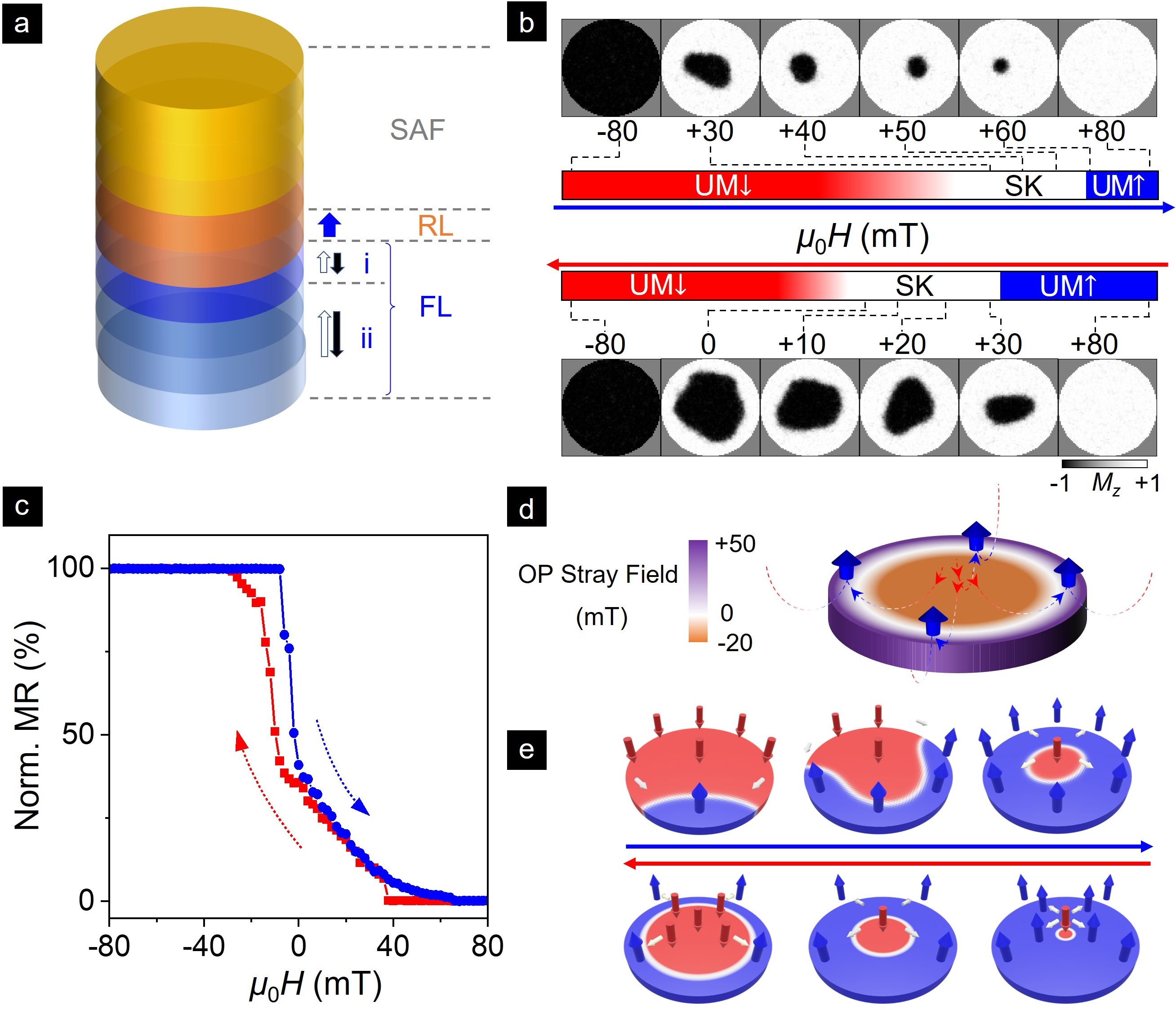} 
	\caption[Simulated Field Evolution]{\textbf{Simulated Field Evolution.}
		\textbf{(a)} Schematic of simulated SK-MTJ stack structure ($W_{\rm Dot} =300$~nm, see Methods).
		\textbf{(b)} Simulated evolution of OP magnetization, $M_{\rm Z}$, of FL for up-sweep (top) and down-sweep (bottom), each showing a distinct reversal mechanism. Centre: colour scalebar indicates statistically determined magnetization states (c.f. \ref{fig:TMR-Expts}c). 
		\textbf{(c)} Simulated normalized MR$(H)$ loop, calculated using a parallel-resistor model (see Methods). 
		\textbf{(d)} Colour plot of spatial distribution of stray field at the FL due to overlying RL-SAF structure, with sizable radial variations near the dot edge. \textbf{(e)} Schematic depiction of skyrmion nucleation process in the FL for up-sweep (top: only for finite stray field) and down-sweep (bottom) respectively.}
	\label{fig:TMR-sims}
\end{figure}

\paragraph{TMR: MFM & Discussion}
To elucidate these peculiar MR characteristics, we used MFM to image the microscopic field evolution of SK-MTJ nanodot arrays ($W_{\rm Dot} \simeq 300$ nm, 24 dots). 
The MFM data, acquired using similar field protocols as the MR, reflects the microscopic evolution of the FL, whose contrast is reduced by intermediate RL and SAF layers. 
\ref{fig:TMR-Expts}c shows the evolution of 3 representative dots, whose FL exhibits 3 distinct states: UM$\uparrow$ (i.e., P, 2c: right), UM$\downarrow$ (i.e., AP, 2c: left) and SK (2c: centre). 
The down-sweep from UM$\uparrow$/P-state (\ref{fig:TMR-Expts}c: bottom) begins reversal at $\sim 30$\,mT by nucleating a down-polarized skyrmion \cite{Boulle.2016, Zeissler.2017, Ho.2019}. 
As $H$ is further reduced, the skyrmion increases in size, eventually forming the UM$\downarrow$/AP-state at negative fields. 
Meanwhile, the up-sweep from UM$\downarrow$/AP-state remains uniform till above ZF (\ref{fig:TMR-Expts}c: top), and suddenly shrinks into a down-polarized skyrmion at $\sim$10-20~mT. 
With increasing $H$, the skyrmion further shrinks and disappears into the UM$\uparrow$/P-state. 
Overall, SK-MTJ dots exhibit asymmetric field evolution, skewed towards positive fields, and with nucleated skyrmions having unique (down) polarity for both sweeps. 
Such asymmetry is unreported for MTJs or skyrmions, and contrasts with the symmetric field evolution of parent films (SM §S4), and in comparable FL dots (SM §S5). 
The underlying nucleation mechanisms will be examined subsequently (\ref{fig:TMR-sims}). 

\paragraph{TMR - Imaging Comparison}
Next, we investigate quantitative congruence between MR and magnetic states in \ref{fig:TMR-Expts}d-e by comparing the normalized MR from $R(H)$ measurements, 
\begin{equation}\label{Eq:Norm-MR} 
{\rm MR}(H) = \left. (R(H)-R_{\rm P}) \right/ (R_{\rm AP}-R_{\rm P})\quad,
\end{equation}
with the MR estimated from the MFM-imaged magnetization configuration (\ref{Eq:SKMTJ-NormMR}, see Methods). 
Both the up-sweep (\ref{fig:TMR-Expts}d) and down-sweep (\ref{fig:TMR-Expts}e) show reasonable agreement between techniques, including positive skew and asymmetry in jumps to/from the AP-state vs. P-state. 
Note that the MFM data is consistently offset by $\sim +10$\,mT (SM §S9), likely due to the flux trapped within its superconducting magnet. 
We conclude that the asymmetry in MR jumps arises from differences in nucleated skyrmion sizes between the up-sweep and down-sweep. 
Reversing the RL orientation gives near-identical asymmetry across both measurements, albeit with a mirrored ${\rm MR}(H)$ loop (SM §S8) and opposite polarity for imaged skyrmions (SM §S5). 
Similar asymmetries in $R(H)$ loops and imaged textures are also observed for varying $W_{\rm Cell} \simeq 250$ and 350 nm (SM §S10), albeit with confinement effects systematically influencing quantitative trends. 
Overall, the observed consistency between MR and imaged states cements our MTJ as an ambient quantitative probe of skyrmions. 

\paragraph{Micromag Simulations}
To elucidate the mechanism underlying the switching asymmetries, we performed grain-free micromagnetic simulations for SK-MTJ and FL stack structures. 
Simulated stacks were designed to reproduce key observations using experiment-derived parameters (\ref{fig:TMR-sims}a, SM §S7), and established recipes were used to simulate hysteresis loops (Methods)\cite{Chen.2022}. 
Throughout the simulations, the composite stack (\ref{fig:TMR-sims}a: i,ii) evolved together as one FL, further validating our experimental interpretation. \ref{fig:TMR-sims}b-c show the microscopic OP magnetization evolution and resulting normalized MR for RL orientation consistent with experiments. 
Magnetization reversal for the down-sweep (\ref{fig:TMR-sims}b: bottom) begins at the dot centre by nucleating a skyrmion of the opposite polarity, and progresses with its expansion to the edge. 
Conversely, for the up-sweep (\ref{fig:TMR-sims}b: top) reversal begins at the dot edge, entrapping a skyrmion at the centre, and progresses with its shrinking and disappearance. 
The asymmetries in microscopic field evolution, which result in a skewed ${\rm MR}(H)$ plot (\ref{fig:TMR-sims}c) agree well with experiments (\ref{fig:TMR-Expts}d-e), and contrasts with the symmetric field evolution of the FL (SM §S7). 

\paragraph{Stray Field \& Nucleation Mechanisms}
The simulations shed valuable light on the stray field origin of the field evolution asymmetries. 
While the SAF overlayer largely compensates for the stray field effect of RL on the FL, such effects are known to re-emerge for smaller MTJ sizes \cite{Han.2015}. 
For our MTJ dots, the OP stray field has a distinctive radial profile (\ref{fig:TMR-sims}d) -- small, negative ($-10$ mT) at the centre, and large, positive ($+50$ mT) at the edge \cite{Zhang.2018}. 
For the up-sweep (from UM$\downarrow$, \ref{fig:TMR-sims}e: top), down-polarization is therefore stabilized at the centre, while switching is initiated at the edge, and delayed. 
In comparison, SK-MTJ films and FL dots do not exhibit this distinctive nucleation mechanism due to the absence of confinement-induced stray fields. 
Conversely, the down-sweep (from UM$\uparrow$, \ref{fig:TMR-sims}e: bottom) aligns with the stray field, and reversal is thus initiated at the centre, as typically expected. 
The asymmetric nucleation mechanisms and switching fields induce a sizable discrepancy in skyrmion sizes, with notable consequences. 

\section{Zero Field Stability\label{sec:Zero-Field}}
\begin{figure}
    \centering \includegraphics[width=1\linewidth]{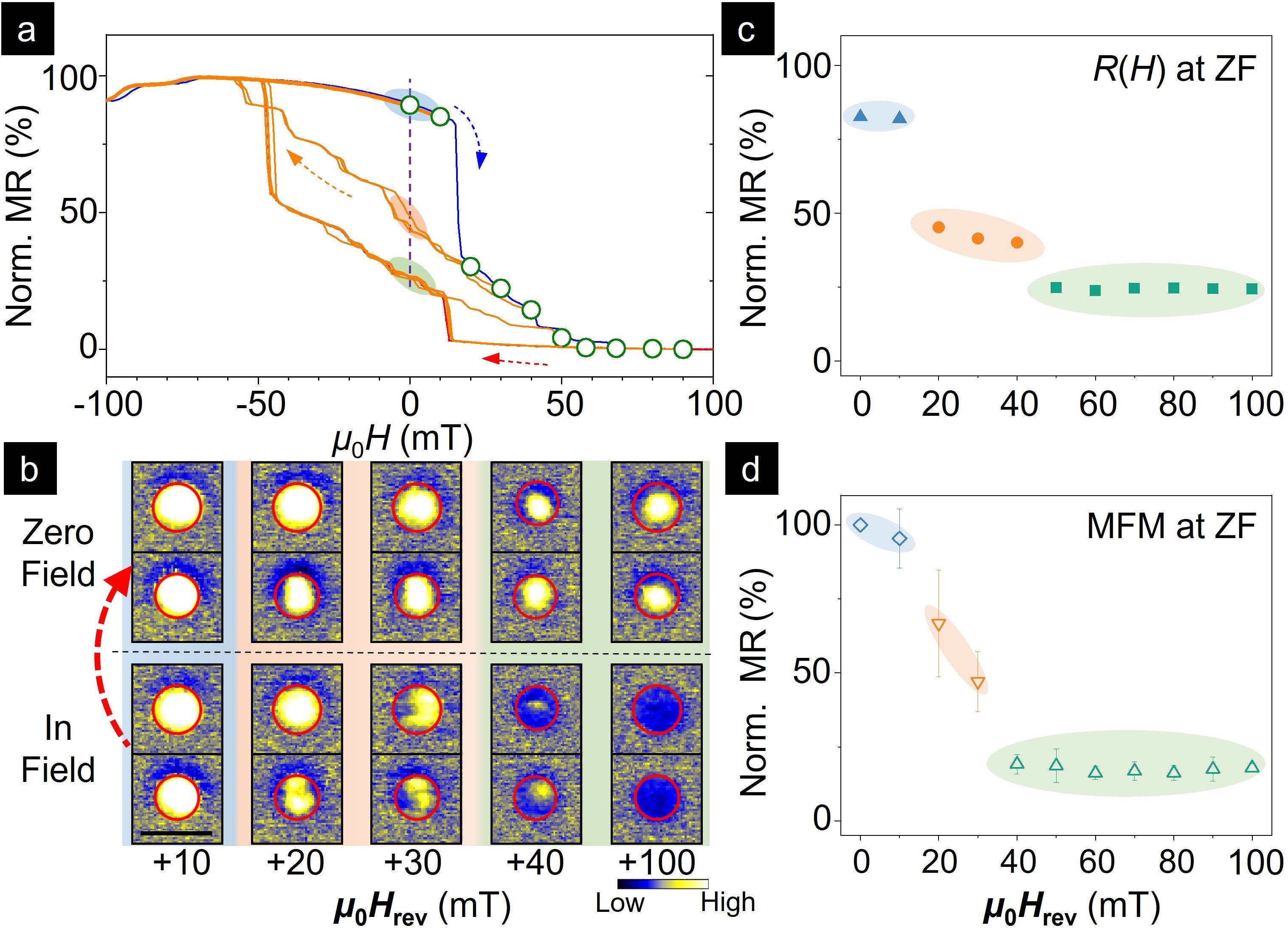}
    \caption[Zero Field Stability]{\textbf{Zero Field Stability.} 
    \textbf{(a)} MR first-order reversal curve loops (MR-FORCs) acquired with varying reversal fields, $H_{\rm rev}$ (open circles), on $W_{\rm Cell} \simeq 300$~nm device. Colours distinguish reversal trends for low (green), intermediate (orange) and high (blue) $H_{\rm rev}$. 
    \textbf{(b)} MFM images of 2 SK-MTJ dots (of 16) acquired in-field, i.e., at $H_{\rm rev}$ (bottom), and then at ZF (top), for varying $H_{\rm rev}$.
    \textbf{(c-d)} Evolution of normalized MR at zero field (ZF, in \%) 
    with $H_{\rm rev}$, obtained from (c) $R(H)$ measurements (in a) and (d) MFM-imaged median magnetic states (in b, \ref{Eq:SKMTJ-NormMR}) 
    Shaded circles in (c-d) distinguish the three ZF MR regimes.}
    \label{fig:ZF-FORC}
\end{figure}

\paragraph{ZF: MR-FORC \& MFM}
Next, we study the ZF stability of skyrmionic MTJ states, by employing first-order reversal curve magnetometry (FORC), used extensively to characterize irreversible magnetization processes\cite{Davies.2004} and skyrmion formation mechanisms\cite{Tan.2020}. 
We performed MR-FORC\cite{Pomeroy.2009} by acquiring a set of minor loops from negative saturation, $-H_{\rm s}$, up to reversal fields, $H_{\rm rev}$, and back. 
With varying $H_{\rm rev}$, the MR($H$) loops for $W_{\rm Cell} \simeq 300$ nm (\ref{fig:ZF-FORC}a) exhibit 3 distinct reversal trends. 
For large $H_{\rm rev}$ ($\geq50$ mT), showing full reversal, and small $H_{\rm rev}$ ($\leq10$ mT), showing no reversal, they closely follow the respective major loop. 
However, for $H_{\rm rev}\simeq 20\text{-}40$ mT, they progress through a distinct intermediate state. 
We acquired MFM data using similar minor loop recipes to visualize the $H_{\rm rev}$ evolution of in-field (i.e., at $H_{\rm rev}$) and ZF states\cite{Tan.2020}. 
\ref{fig:ZF-FORC}b shows that reversals from all but the smallest $H_{\rm rev}$ values result in ZF skyrmions, albeit with observable size variations. 

\paragraph{ZF States: Discussion}
In analogy with \ref{fig:TMR-Expts}d-e, we compare the $H_{\rm rev}$ evolution of normalized MR at ZF obtained from MR-FORC (\ref{fig:ZF-FORC}c), with that from MFM-imaged magnetic states (\ref{fig:ZF-FORC}d). 
The two MR trends consistently show 3 distinct ZF states stabilized with varying $H_{\rm rev}$, with good quantitative agreement. 
Notably, both identify, in addition to the two major loop ZF states, i.e., UM$\downarrow$ (${\rm MR}\gtrsim 90$\%) and SK (${\rm MR}\sim 20-30$\%), an intermediate ZF state (${\rm MR} \sim 50$\%).
Closer inspection reveals that this intermediate ZF state, formed for $H_{\rm rev} \sim$ 20-30 mT, is a larger skyrmion, distinct from the major loop SK state. 
This metastable state, which arises from field-reversal of larger skyrmions (\ref{fig:ZF-FORC}b: bottom), may be stabilized by the inhomogeneous stray field profile trapping the skyrmion edge, and pinning it to the up-magnetized ZF state. 
The nonvolatility of these distinct MTJ states lends itself to future explorations of multi-bit functionalities. 

\section{Electrical Manipulation\label{sec:E-Manip}}

\begin{figure}
    \centering \includegraphics[width=1\linewidth]{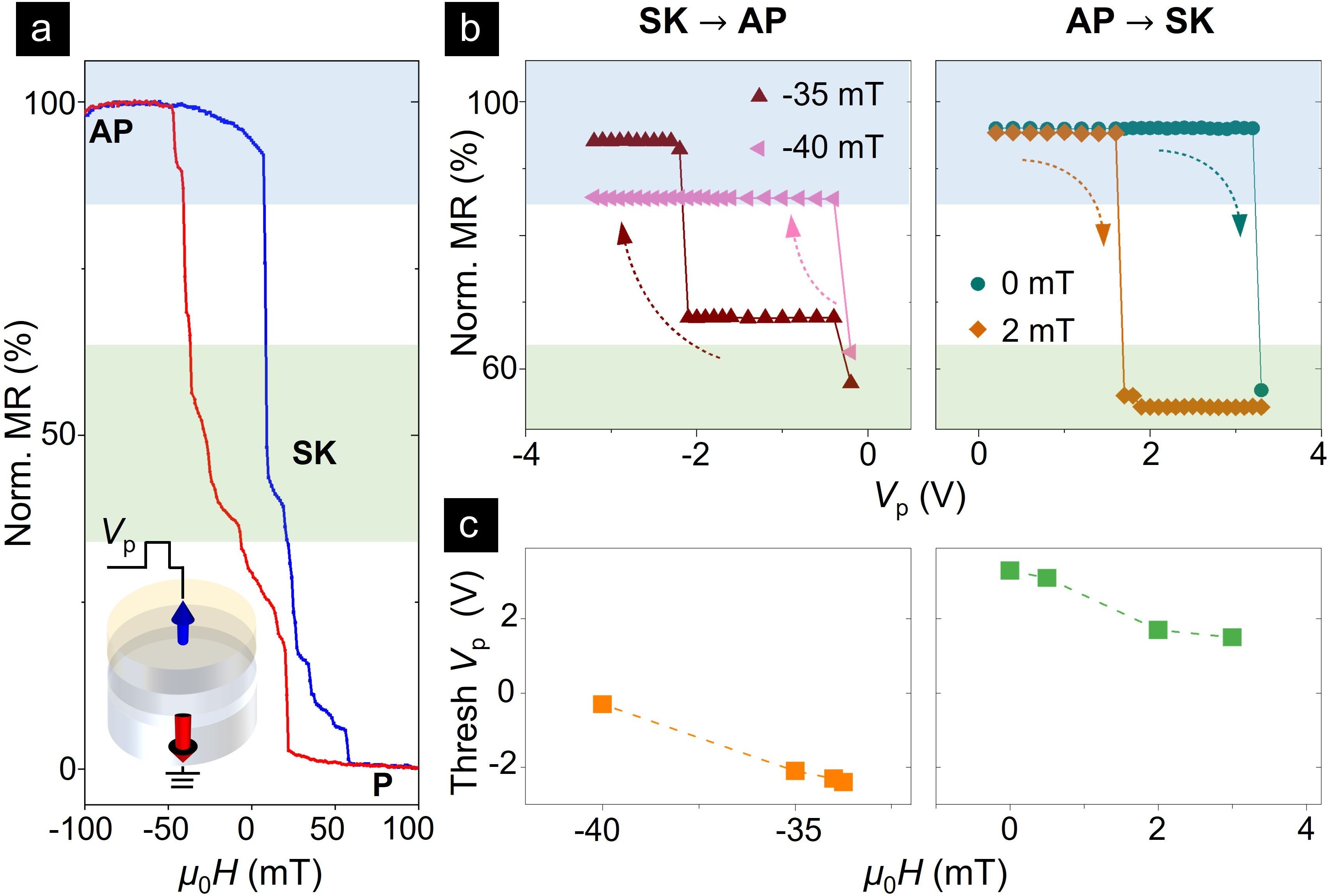}
    \caption[Electrical Switching]{\textbf{Electrical Switching.} 
    \textbf{(a)} MR$(H)$ loop of SK-MTJ used for electrical switching experiments, with MR regimes of FL states (AP: blue, SK: green, P) indicated. Bottom inset: schematic of voltage pulse $V_{\rm p}$ (width: 10~ns) applied to device. 
    \textbf{(b)} Pulsed switching from SK $\rightarrow$ AP (left) and from AP $\rightarrow$ SK (right) states, realized with increasing $\left|V_{\rm p}\right|$ (arrows) for different $H$. Switching transitions, identified by MR changes, occur above a threshold $\left|V_{\rm p}\right|$ for each field. 
    \textbf{(c)} Variation of threshold $V_{\rm p}$ with $H$ for SK $\rightarrow$ AP (left) and AP $\rightarrow$ SK (right) transitions.}
    \label{fig:E-Manip}
\end{figure}

\paragraph{Electrical Switching: Methods \& Results}
A critical requirement for functional devices is the electrical manipulation of resistive states. 
Accordingly, we characterize the response of SK-MTJs ($W_{\rm Cell} \simeq 300$ nm) to voltage pulses of amplitude $V_{\rm p}$ (width $\sim 10$ ns) applied across the MTJ. 
An OP field was applied to initialize the device, and the measured MR was used to determine the initial and final states (see \ref{fig:E-Manip}a). 
Notably, when initialized in the AP state (\ref{fig:E-Manip}b: right), ZF switching to the SK state is observed for positive $V_{\rm p}$ ($\geq 3.2$ V). 
Such AP$\rightarrow$SK switching, or electrical skyrmion nucleation, can also be achieved for small, positive $H$. 
While the threshold switching voltage  for $H>0$ is reduced c.f. ZF (\ref{fig:E-Manip}c: right), the final MR is similar (\ref{fig:E-Manip}b: right), indicating negligible variation in nucleated skyrmion size. 
Conversely, skyrmion deletion, i.e., switching from SK to AP state, can be achieved using negative $V_{\rm p}$ ($\leq -2$ V), and at negative $\mu_0 H$ (\ref{fig:E-Manip}b: left). 
Here as well, reducing $\left|H\right|$ increases the magnitude of threshold voltage required for skyrmion deletion (\ref{fig:E-Manip}c: left). 

\paragraph{Electrical Switching: Mechanisms}
Remarkably, the threshold current density for AP$\leftrightarrow$SK switching ($4.8\times10^7$\,A/m$^2$), is 2-3 orders of magnitude lower than that required for spin transfer torque (STT $\sim 10^{10}$\,A/m$^2$) \cite{Ikeda.2010,Watanabe.2018} or spin-orbit torque ($\sim 10^9$\,A/m$^2$) \cite{Liu.2012, Wang.2018, Cubukcu.2018} driven switching of P-MTJs at similar pulse widths. 
In fact, the switching current and energy ($\sim 100$~fJ) are also 100-1000 times lower than values reported for nucleating and deleting skyrmions, including at finite fields, and across a variety of geometries and mechanisms \cite{Buttner.2017, Woo.2018, Finizio.2019, Bhattacharya.2020, Li.2022}. 
This leads us to examine alternative mechanisms that may assist STT switching of skyrmions in MTJs, notable among which is voltage control of magnetic anisotropy (VCMA) at the CoFeB/MgO interface \cite{Wang.2012, Niranjan.2010}. 
Indeed, we find that applying a positive (negative) DC voltage across the MTJ can decrease (increase) the MR-measured anisotropy (SM §S11). The estimated VCMA coefficient, -41.5\,fJ/V$\cdot$m, is in line with reports on CoFeB/MgO interfaces \cite{Wang.2012, Li.2017}. 
Such voltage-induced decrease (increase) of anisotropy, acting as a pseudo-magnetic field, is known to induce the nucleation (deletion) of skyrmions \cite{Hsu.2017, Schott.2017, Bhattacharya.2020}. 
In our MTJ devices, a combination of VCMA and STT effects may drive skyrmion nucleation and deletion, while defects and interlayer coupling may stabilize the final state \cite{Buttner.2017, Zhang.2022}. 
Forthcoming efforts could establish details of the underlying mechanism, while optimized stack design, device geometry, and pulsing recipes may enable ZF switching between two or more MTJ states. 

\section{Outlook}

\paragraph{Summary of Results}
We have realized nanoscale, two-terminal chiral MTJs hosting a single Néel skyrmion in addition to uniform P/AP states.
The presence of the skyrmion, and its size, can be quantitatively determined via its sizable $\sim 20-70$\% MR readout. 
Crucially, the distinctive stray field profile of the nanoscale MTJ ensures that formed skyrmions have fixed polarity, nucleated via two distinct mechanisms.
This, in turn, enables the ZF stability of skyrmions of distinct sizes, relevant to multi-state functionality. 
Finally, skyrmions can be electrically nucleated and deleted in the MTJ using current densities 2-3 orders lower than state-of-the-art.

\paragraph{Skyrmionic Racetrack Devices}
The quantitative electrical readout of skyrmions established herein is a much-anticipated milestone for the burgeoning field of skyrmionics. 
While ease of lateral skyrmion manipulation is touted for numerous racetrack-like device applications, the lack of a robust readout technique has been a longstanding roadblock. 
The realization of MTJ detection of skyrmions within stacks compatible with lateral manipulation promises their imminent integration with existing skyrmionic device architectures. 
Such MTJs may be used to detect or count skyrmions, and to quantify their size – forming a platform for a wide range of binary, multi-bit, and/or analogue applications. 
Meanwhile, the presented ease of skyrmion generation and deletion with unprecedented efficiency enhances their utility for logic and unconventional computing. 

\paragraph{Multi-bit MTJs \& New Concepts}
Concomitantly, our work – realized on a 200 mm fabrication platform – also presents a lucrative pathway for the development of next-generation MTJs. 
In order to overcome existing limitations, efforts on MTJs have explored complex fabrication geometries, as well as multi-terminal and multi-device architectures. 
In contrast, skyrmions provide a facile approach to realize multi-state functionality within established two-terminal architecture, with potential to improve scalability and efficiency. 
Therefore, we expect that skyrmionic MTJs may imminently be the subject of integrated circuit-level studies, with potential for rapid development and application relevance. 
Moreover, the ubiquity of skyrmions within complex materials and heterostructures hosting novel magnetism, superconductivity, and beyond, may spawn hybrid tunnel junction devices with applications in biomimetic and quantum computing. 
\begin{center}\rule[0.5ex]{0.5\columnwidth}{0.5pt}\par\end{center}
\section{Methods \label{sec:Methods}}
\setcounter{paragraph}{0}
\setlength{\parskip}{0.2ex plus0.2ex minus0.2ex}
\titlespacing*{\paragraph}{0em}{0ex}{0em}[]

\begin{small}
\paragraph{Film Deposition}
\textsf{\textbf{Multilayer Thin Films}} corresponding to full (SK,P)-MTJ, companion MTJ, and FL stack structures were sputter-deposited at RT on thermally oxidized 200 mm Si/SiO$_{2}$ wafers using the Singulus Timaris\texttrademark\, ultrahigh vacuum physical vapour deposition system. 
The full MTJ stack (\ref{fig:Device}a) was deposited with the following functional layers (thicknesses in nm): 
\begin{enumerate}[nosep, leftmargin=*]
\item Bottom electrode: Ta(5)/TaN(44)/[Ta(5)/ Ru(7)]$_{2}$,
\item Buffer layer: TaN(5)/Pt(5),
\item FL (ii): [Ir(1)/Fe(0.2)/Co($x$:0.8-1.6)/Pt(1)]$_{3}$,
\item FL (i): W(2)/Co$_{20}$Fe$_{60}$B$_{20}$($y$:0.9-1.2),
\item Tunnel barrier: MgO(2.2),
\item RL: Co$_{20}$Fe$_{60}$B$_{20}$(1.3),
\item SAF: W(0.6)/Co(0.6)/Pt(1.5)/[Co(0.4)/Pt(0.2)]$_4$/Co(0.5) /Ru(1.0)/Co(0.5)/[Pt(0.2)/Co(0.4)]$_{13}$,
\end{enumerate}
and capped with a sacrificial overlayer.
The SK-MTJ corresponds to $x/y \equiv 1.3/1.1$, and the P-MTJ to $0.9/0.9$. 
Companion SK-MTJ (no bottom electrode, overlayer) and FL (no bottom electrode, RL, SAF, overlayer) stacks were fabricated for comparative purposes (see SM §S1), including magnetic imaging. Post deposition, all wafers were annealed \emph{in situ} at 300$^\circ$C for 60 min, and cooled naturally.

\paragraph{Device Fabrication}
\textsf{\textbf{Device Fabrication.}} Full MTJ stack wafers were patterned into devices ($W_{\rm Cell}$: 200–1000 nm) using a dedicated deep-UV photomask at A{*}STAR’s 200 mm fabrication facility. 
Major steps included lithography using Canon EX5\texttrademark\, stepper, Ar-ion beam etching (Oxford CAIBE\texttrademark), oxide filling for MTJ protection, followed by planarization.
Finally, top electrodes Ta(10)/Al(180) were patterned for electrical connections. 
Specific etch conditions were developed to prevent sidewall shorting and mitigate MTJ skirting. 
Separately, companion SK-MTJ and FL stacks were patterned into dots ($W_{\rm Dot}$: 250–500 nm) using electron beam lithography (Elionix\texttrademark) and ion-beam etching (Oxford CAIBE\texttrademark).

\paragraph{Structural \& Compositional Imaging}
\textsf{\textbf{Structural Characterization.}} The plan view of the $W_{\rm Cell} \simeq$ 300 nm SK-MTJ device was imaged by a FEI Helios\texttrademark\, Nanolab 600 focused ion beam scanning electron microscope. 
Subsequently, a cross-section lamella of the device was prepared, and a FEI Tecnai\texttrademark\, F20XT TEM, equipped with an energy dispersive X-ray spectroscopy functionality was used for cross-sectional view imaging and elemental analysis.  

\paragraph{Quantifying Magnetic Properties}
\textsf{\textbf{Magnetic Properties.}} Magnetization measurements were performed on diced film coupons using the Princeton\texttrademark\, alternating gradient magnetometer. 
For the FL within SK-MTJ (1.3/1.1) and P-MTJ (0.9/0.9) stacks, the saturation magnetization, $M_{\rm s}$, was measured to be $\sim$1.12\,MA/m, and the effective anisotropy, $K_{\rm eff}$,  is 0.03\,MJ/m$^3$ and 0.20\,MJ/m$^3$, respectively (SM §S3). 

Wafer-level electrical measurements were performed using a CAPRES CIPTech-M300\texttrademark\, CIP-TMR system.
For the MTJ wafer, the measured TMR, $R_{\rm AP}/R_{\rm P} - 1$, was $0.97 \pm 0.02$ (SM §S1), and resistance-area product was $23 \pm 5$\,k$\Omega\,\mu$m$^2$. Here, error bars represent variation across wafer.

The Dzyaloshinskii-Moriya interaction of the FL components -- [Ir/Fe(0.2)/Co(1.3)/Pt]$_3$: $D \simeq -1.28$\,mJ/m$^2$; W/CFB(1.1)/MgO: $D \simeq 0.05$\,mJ/m$^2$ -- were determined via spin wave spectroscopy by performing Brillouin light scattering in Damon-Eshbach geometry (SM §S3). 
Meanwhile, the interlayer exchange coupling (IEC) between the two FL components, $A_{\rm IEC} \simeq 0.93$ mJ/m$^2$, was estimated using a mixed anisotropy magnetometry technique on a customized control stack (SM §S3), similar to previous works \cite{Chen.2023}. The IEC trend was validated over a range of spacer thicknesses.

\paragraph{Lorentz TEM Imaging}
\textbf{\textsf{Lorentz TEM Measurements}} were performed using an FEI Titan\texttrademark\, S/TEM on companion SK-MTJ and FL stacks deposited onto 20 nm thick SiO$_2$ membranes. 
The TEM was operated in Fresnel mode at 300 kV, with a defocus of $-2.4$~mm, and the objective lens was used to apply \emph{in situ} OP magnetic fields.
A tilt of 15$^{\circ}$ with respect to normal incidence was used to ensure sufficient contrast from Néel textures. 
Custom-written Python scripts were used to perform background subtraction. 

\paragraph{MFM Imaging}
\textbf{\textsf{MFM Measurements}} were performed using an AFM-in-cryostat setup from Attocube\texttrademark\, systems using Co-alloy coated SSS-MFMR\texttrademark\, tips (diameter $\sim$30 nm, moment $\sim$80 kA/m) by NanoSensors Inc. 
SSS tips were chosen to ensure high-resolution imaging with minimal perturbation of the FL (lift height $\sim15$~nm, intermediary layers $\sim30$~nm).
Following OP saturation at $-1.5$ T, the evolution of SK-MTJ nanodots was imaged over OP fields of $\pm100$~mT. 
A total of 126 SK-MTJ dots of varying sizes ($W_{\rm Dot}$: 250-500 nm, companion MTJ wafer) were imaged, of which 24 dots corresponded to $W_{\rm Dot}\simeq300$ nm. 
Comparative imaging was also performed for RL saturated with opposite polarity ($+1.5$\,T), and on FL dots (i.e., no RL, SAF). 

\paragraph{MFM Analysis} 
\textbf{\textsf{MFM Analysis.}} All MFM images presented in the manuscript are referenced to positive tip magnetization. 
To account for flipping of tip magnetization, the phase contrast of MFM images acquired with negative tip polarity ($H \lesssim 25$ mT) were inverted (SM §S6). 
The sizes of imaged skyrmions were determined from their full width at half maximum of their fitted 2D anisotropic Gaussian function (SM §S6). 
Quantitative one-to-one correspondence between the size, $d_{\rm S,fit}$ obtained from the Gaussian fit to MFM data and the expected profile of a Néel skyrmion with diameter $d_{\rm S}$ was established using micromagnetic simulations (SM §S7).

For each dot and field, the MFM-imaged state, characterized by its areal magnetization fraction, $f_{\rm M} \in \{0,1\}$, was classified as UM$\uparrow$ ($f_{\rm M} = 0$), UM$\downarrow$ ($f_{\rm M} = 1$), or SK ($f_{\rm M} = d^2_{\rm S}/W^2_{\rm Dot}$). 
Here, $d_{\rm S}$ ($\equiv d_{\rm S,fit}$), is referenced to $W_{\rm Dot} \simeq 290$~nm, which corresponds to the MFM-imaged size of UM$\uparrow,\downarrow$ states averaged across all dots.
A clear distinction between SK and UM states was established using the minimum threshold for amplitude of the fitted Gaussian function. 
The MFM-estimated MR in \ref{fig:TMR-Expts}d-e (\ref{fig:ZF-FORC}d) utilizes the median magnetization state, $\overline{f_{\rm M}}(H)$, of 3 representative dots out of 24 (16) imaged.

\paragraph{Electrical Measurement Setup}
\textbf{\textsf{Electrical $R(H)$ Measurements}} were performed on a custom-designed probe station, with OP magnetic fields of up to $\pm$600 mT.
MR was measured using AC lock-in technique (source voltage: 4 mV, frequency: 317 Hz, \ref{fig:Device}d), by determining the voltage drop, $V_{\rm AC}$, across the reference resistor and device under test (see SM §S8). 
Electrical switching experiments used a Tektronix AFG3251\texttrademark\, pulse generator, while bias voltages were applied using a Yokogawa GS210\texttrademark\, voltage source, with pulse width set to 10 ns, and delay to 8 s (see SM §S11). 

\paragraph{MR of SK-MTJ}
\textbf{\textsf{MR of SK-MTJ.}} Here, we describe the procedure used to estimate the MR of the SK-MTJ from its imaged magnetization state. 
Skyrmions are formed with AP polarity relative to P background. A skyrmion should form an AP conductance channel proportional to its areal fraction, $f_{\rm M} = d^2_{\rm S}/W^2_{\rm Dot}$, where the domain wall contribution is assumed to be negligible. 
The parallel resistor model \cite{Chen.2022b} then gives the MTJ resistance as
\begin{align*} 
    R_{\rm MTJ}^{-1}(H) 
        &=  \left[R_{\rm AP} / f_{\rm M}(H)\right]^{-1} 
            + \left[(R_{\rm P} / (1 - f_{\rm M}(H))\right]^{-1} \\
\implies R_{\rm MTJ}(H) &= \frac{R_{\rm AP}\cdot R_{\rm P}}{R_{\rm AP}
            +f_{\rm M}(H)\cdot\left(R_{\rm P}-R_{\rm AP}\right)}\quad,\\
\end{align*}
The normalized MR, defined in \ref{Eq:Norm-MR}, is therefore given as
\begin{equation}\label{Eq:SKMTJ-TMR-2}
{\rm MR}(H) = \frac{R_{\rm P}\cdot f_{\rm M}(H)}{R_{\rm AP}\left(1 - f_{\rm M}(H)\right) 
            + R_{\rm P}\cdot f_{\rm M}(H)}
\end{equation}
\ref{Eq:SKMTJ-TMR-2} reduces to the expression in \ref{Eq:SKMTJ-NormMR}, which holds across all magnetization states, $f_{\rm M} \in \{0,1\}$ realized in this work.
The MR of the imaged states is then estimated by using the empirically measured $\overline{f_{\rm M}}(H)$, as detailed above. 

\paragraph{Micromagnetic Simulations \& Analysis}
\textbf{\textsf{Micromagnetic Simulations}} were performed using mumax$^3$ \cite{Vansteenkiste.2014}. 
A  cylindrical geometry was defined within a simulated box of 300 $\times$ 300 $\times$ 31.5 nm, with cell sizes of 4.2 nm $\times$ 4.2 nm $\times$ 3.5 nm. 
The FL stacks were represented using an effective medium model \cite{Woo.2016}, while accounting for interlayer coupling. 
The parameters used for simulations were consistent with those obtained from experimental measurements (see SM §S7).
Hysteresis loops were simulated using established protocols described in previous works \cite{Vansteenkiste.2014, Chen.2022}. 
Stray field analysis was performed by setting the magnetization of FL stack to zero, and then calculating the stray field of FL interlayer. 

The simulated TMR was calculated using a parallel resistor model, with the magnetoresistivity, $\rho_{\rm M}(H)$, defined as 
\begin{equation*}
\rho_{\rm M}(H) = \left.\left(1-\cos\theta(H))(\rho_{\rm AP} - \rho_{\rm P}\right) \right/ 2 \,+\, \rho_{\rm P} \quad .
\end{equation*}
Here, $\rho_{\rm AP,P}$ are the areal resistivities of the AP,P states, while $\theta(H)$ is the angle of relative magnetization between the FL and RL respectively.
The MTJ resistance is then given by
\begin{equation*}
R_{\rm MTJ}^{-1}(H) =  \mathcal{C}\,\int{dS\cdot\rho_{\rm M}^{-1}(H)}
\end{equation*}
where the constant $\mathcal{C}$ is related to the device geometry.
\end{small}

\phantomsection
\def\bibsection{\section*{\refname}} 
\linespread{1.00}
\setlength{\parskip}{0.2ex}
\bibliographystyle{apsrev4-1}
\bibliography{SK-MTJ_Refs}

\begin{thebibliography}{51}%
\makeatletter
\providecommand \@ifxundefined [1]{%
 \@ifx{#1\undefined}
}%
\providecommand \@ifnum [1]{%
 \ifnum #1\expandafter \@firstoftwo
 \else \expandafter \@secondoftwo
 \fi
}%
\providecommand \@ifx [1]{%
 \ifx #1\expandafter \@firstoftwo
 \else \expandafter \@secondoftwo
 \fi
}%
\providecommand \natexlab [1]{#1}%
\providecommand \enquote  [1]{``#1''}%
\providecommand \bibnamefont  [1]{#1}%
\providecommand \bibfnamefont [1]{#1}%
\providecommand \citenamefont [1]{#1}%
\providecommand \href@noop [0]{\@secondoftwo}%
\providecommand \href [0]{\begingroup \@sanitize@url \@href}%
\providecommand \@href[1]{\@@startlink{#1}\@@href}%
\providecommand \@@href[1]{\endgroup#1\@@endlink}%
\providecommand \@sanitize@url [0]{\catcode `\\12\catcode `\$12\catcode
  `\&12\catcode `\#12\catcode `\^12\catcode `\_12\catcode `\%12\relax}%
\providecommand \@@startlink[1]{}%
\providecommand \@@endlink[0]{}%
\providecommand \url  [0]{\begingroup\@sanitize@url \@url }%
\providecommand \@url [1]{\endgroup\@href {#1}{\urlprefix }}%
\providecommand \urlprefix  [0]{URL }%
\providecommand \Eprint [0]{\href }%
\providecommand \doibase [0]{http://dx.doi.org/}%
\providecommand \selectlanguage [0]{\@gobble}%
\providecommand \bibinfo  [0]{\@secondoftwo}%
\providecommand \bibfield  [0]{\@secondoftwo}%
\providecommand \translation [1]{[#1]}%
\providecommand \BibitemOpen [0]{}%
\providecommand \bibitemStop [0]{}%
\providecommand \bibitemNoStop [0]{.\EOS\space}%
\providecommand \EOS [0]{\spacefactor3000\relax}%
\providecommand \BibitemShut  [1]{\csname bibitem#1\endcsname}%
\let\auto@bib@innerbib\@empty
\bibitem [{\citenamefont {Julliere}(1975)}]{Julliere.1975}%
  \BibitemOpen
  \bibfield  {author} {\bibinfo {author} {\bibfnamefont {M.}~\bibnamefont
  {Julliere}},\ }\href {\doibase 10.1016/0375-9601(75)90174-7} {\bibfield
  {journal} {\bibinfo  {journal} {Physics Letters A}\ }\textbf {\bibinfo
  {volume} {54}},\ \bibinfo {pages} {225} (\bibinfo {year} {1975})}\BibitemShut
  {NoStop}%
\bibitem [{\citenamefont {Moodera}\ \emph {et~al.}(1995)\citenamefont
  {Moodera}, \citenamefont {Kinder}, \citenamefont {Wong},\ and\ \citenamefont
  {Meservey}}]{Moodera.1995}%
  \BibitemOpen
  \bibfield  {author} {\bibinfo {author} {\bibfnamefont {J.~S.}\ \bibnamefont
  {Moodera}}, \bibinfo {author} {\bibfnamefont {L.~R.}\ \bibnamefont {Kinder}},
  \bibinfo {author} {\bibfnamefont {T.~M.}\ \bibnamefont {Wong}}, \ and\
  \bibinfo {author} {\bibfnamefont {R.}~\bibnamefont {Meservey}},\ }\href
  {\doibase 10.1103/PhysRevLett.74.3273} {\bibfield  {journal} {\bibinfo
  {journal} {Physical Review Letters}\ }\textbf {\bibinfo {volume} {74}},\
  \bibinfo {pages} {3273} (\bibinfo {year} {1995})}\BibitemShut {NoStop}%
\bibitem [{\citenamefont {Slonczewski}(1996)}]{Slonczewski.1996}%
  \BibitemOpen
  \bibfield  {author} {\bibinfo {author} {\bibfnamefont {J.~C.}\ \bibnamefont
  {Slonczewski}},\ }\href {\doibase
  https://doi.org/10.1016/0304-8853(96)00062-5} {\bibfield  {journal} {\bibinfo
   {journal} {Journal of Magnetism and Magnetic Materials}\ }\textbf {\bibinfo
  {volume} {159}},\ \bibinfo {pages} {L1} (\bibinfo {year} {1996})}\BibitemShut
  {NoStop}%
\bibitem [{\citenamefont {Yuasa}\ \emph {et~al.}(2004)\citenamefont {Yuasa},
  \citenamefont {Nagahama}, \citenamefont {Fukushima}, \citenamefont {Suzuki},\
  and\ \citenamefont {Ando}}]{Yuasa.2004}%
  \BibitemOpen
  \bibfield  {author} {\bibinfo {author} {\bibfnamefont {S.}~\bibnamefont
  {Yuasa}}, \bibinfo {author} {\bibfnamefont {T.}~\bibnamefont {Nagahama}},
  \bibinfo {author} {\bibfnamefont {A.}~\bibnamefont {Fukushima}}, \bibinfo
  {author} {\bibfnamefont {Y.}~\bibnamefont {Suzuki}}, \ and\ \bibinfo {author}
  {\bibfnamefont {K.}~\bibnamefont {Ando}},\ }\href {\doibase 10.1038/nmat1257}
  {\bibfield  {journal} {\bibinfo  {journal} {Nature Materials}\ }\textbf
  {\bibinfo {volume} {3}},\ \bibinfo {pages} {868} (\bibinfo {year}
  {2004})}\BibitemShut {NoStop}%
\bibitem [{\citenamefont {Parkin}\ \emph {et~al.}(2004)\citenamefont {Parkin},
  \citenamefont {Kaiser}, \citenamefont {Panchula}, \citenamefont {Rice},
  \citenamefont {Hughes}, \citenamefont {Samant},\ and\ \citenamefont
  {Yang}}]{Parkin.2004}%
  \BibitemOpen
  \bibfield  {author} {\bibinfo {author} {\bibfnamefont {S.~S.~P.}\
  \bibnamefont {Parkin}}, \bibinfo {author} {\bibfnamefont {C.}~\bibnamefont
  {Kaiser}}, \bibinfo {author} {\bibfnamefont {A.}~\bibnamefont {Panchula}},
  \bibinfo {author} {\bibfnamefont {P.~M.}\ \bibnamefont {Rice}}, \bibinfo
  {author} {\bibfnamefont {B.}~\bibnamefont {Hughes}}, \bibinfo {author}
  {\bibfnamefont {M.}~\bibnamefont {Samant}}, \ and\ \bibinfo {author}
  {\bibfnamefont {S.~H.}\ \bibnamefont {Yang}},\ }\href {\doibase
  10.1038/nmat1256} {\bibfield  {journal} {\bibinfo  {journal} {Nature
  Materials}\ }\textbf {\bibinfo {volume} {3}},\ \bibinfo {pages} {862}
  (\bibinfo {year} {2004})}\BibitemShut {NoStop}%
\bibitem [{\citenamefont {Ikeda}\ \emph {et~al.}(2010)\citenamefont {Ikeda},
  \citenamefont {Miura}, \citenamefont {Yamamoto}, \citenamefont {Mizunuma},
  \citenamefont {Gan}, \citenamefont {Endo}, \citenamefont {Kanai},
  \citenamefont {Hayakawa}, \citenamefont {Matsukura},\ and\ \citenamefont
  {Ohno}}]{Ikeda.2010}%
  \BibitemOpen
  \bibfield  {author} {\bibinfo {author} {\bibfnamefont {S.}~\bibnamefont
  {Ikeda}}, \bibinfo {author} {\bibfnamefont {K.}~\bibnamefont {Miura}},
  \bibinfo {author} {\bibfnamefont {H.}~\bibnamefont {Yamamoto}}, \bibinfo
  {author} {\bibfnamefont {K.}~\bibnamefont {Mizunuma}}, \bibinfo {author}
  {\bibfnamefont {H.~D.}\ \bibnamefont {Gan}}, \bibinfo {author} {\bibfnamefont
  {M.}~\bibnamefont {Endo}}, \bibinfo {author} {\bibfnamefont {S.}~\bibnamefont
  {Kanai}}, \bibinfo {author} {\bibfnamefont {J.}~\bibnamefont {Hayakawa}},
  \bibinfo {author} {\bibfnamefont {F.}~\bibnamefont {Matsukura}}, \ and\
  \bibinfo {author} {\bibfnamefont {H.}~\bibnamefont {Ohno}},\ }\href {\doibase
  10.1038/nmat2804} {\bibfield  {journal} {\bibinfo  {journal} {Nature
  Materials}\ }\textbf {\bibinfo {volume} {9}},\ \bibinfo {pages} {721}
  (\bibinfo {year} {2010})}\BibitemShut {NoStop}%
\bibitem [{\citenamefont {Engel}\ \emph {et~al.}(2005)\citenamefont {Engel},
  \citenamefont {Akerman}, \citenamefont {Butcher}, \citenamefont {Dave},
  \citenamefont {DeHerrera}, \citenamefont {Durlam}, \citenamefont
  {Grynkewich}, \citenamefont {Janesky}, \citenamefont {Pietambaram},
  \citenamefont {Rizzo}, \citenamefont {Slaughter}, \citenamefont {Smith},
  \citenamefont {Sun},\ and\ \citenamefont {Tehrani}}]{Engel.2005}%
  \BibitemOpen
  \bibfield  {author} {\bibinfo {author} {\bibfnamefont {B.~N.}\ \bibnamefont
  {Engel}}, \bibinfo {author} {\bibfnamefont {J.}~\bibnamefont {Akerman}},
  \bibinfo {author} {\bibfnamefont {B.}~\bibnamefont {Butcher}}, \bibinfo
  {author} {\bibfnamefont {R.~W.}\ \bibnamefont {Dave}}, \bibinfo {author}
  {\bibfnamefont {M.}~\bibnamefont {DeHerrera}}, \bibinfo {author}
  {\bibfnamefont {M.}~\bibnamefont {Durlam}}, \bibinfo {author} {\bibfnamefont
  {G.}~\bibnamefont {Grynkewich}}, \bibinfo {author} {\bibfnamefont
  {J.}~\bibnamefont {Janesky}}, \bibinfo {author} {\bibfnamefont {S.~V.}\
  \bibnamefont {Pietambaram}}, \bibinfo {author} {\bibfnamefont {N.~D.}\
  \bibnamefont {Rizzo}}, \bibinfo {author} {\bibfnamefont {J.~M.}\ \bibnamefont
  {Slaughter}}, \bibinfo {author} {\bibfnamefont {K.}~\bibnamefont {Smith}},
  \bibinfo {author} {\bibfnamefont {J.~J.}\ \bibnamefont {Sun}}, \ and\
  \bibinfo {author} {\bibfnamefont {S.}~\bibnamefont {Tehrani}},\ }\href
  {\doibase 10.1109/TMAG.2004.840847} {\bibfield  {journal} {\bibinfo
  {journal} {IEEE Transactions on Magnetics}\ }\textbf {\bibinfo {volume}
  {41}},\ \bibinfo {pages} {132} (\bibinfo {year} {2005})}\BibitemShut
  {NoStop}%
\bibitem [{\citenamefont {Chappert}\ \emph {et~al.}(2007)\citenamefont
  {Chappert}, \citenamefont {Fert},\ and\ \citenamefont
  {Van~Dau}}]{Chappert.2007}%
  \BibitemOpen
  \bibfield  {author} {\bibinfo {author} {\bibfnamefont {C.}~\bibnamefont
  {Chappert}}, \bibinfo {author} {\bibfnamefont {A.}~\bibnamefont {Fert}}, \
  and\ \bibinfo {author} {\bibfnamefont {F.~N.}\ \bibnamefont {Van~Dau}},\
  }\href {\doibase 10.1038/nmat2024} {\bibfield  {journal} {\bibinfo  {journal}
  {Nature Materials}\ }\textbf {\bibinfo {volume} {6}},\ \bibinfo {pages} {813}
  (\bibinfo {year} {2007})}\BibitemShut {NoStop}%
\bibitem [{\citenamefont {Dieny}\ \emph {et~al.}(2020)\citenamefont {Dieny},
  \citenamefont {Prejbeanu}, \citenamefont {Garello}, \citenamefont
  {Gambardella}, \citenamefont {Freitas}, \citenamefont {Lehndorff},
  \citenamefont {Raberg}, \citenamefont {Ebels}, \citenamefont {Demokritov},
  \citenamefont {Akerman}, \citenamefont {Deac}, \citenamefont {Pirro},
  \citenamefont {Adelmann}, \citenamefont {Anane}, \citenamefont {Chumak},
  \citenamefont {Hirohata}, \citenamefont {Mangin}, \citenamefont {Valenzuela},
  \citenamefont {Onbaşlı}, \citenamefont {d’Aquino}, \citenamefont
  {Prenat}, \citenamefont {Finocchio}, \citenamefont {Lopez-Diaz},
  \citenamefont {Chantrell}, \citenamefont {Chubykalo-Fesenko},\ and\
  \citenamefont {Bortolotti}}]{Dieny.2020}%
  \BibitemOpen
  \bibfield  {author} {\bibinfo {author} {\bibfnamefont {B.}~\bibnamefont
  {Dieny}}, \bibinfo {author} {\bibfnamefont {I.~L.}\ \bibnamefont
  {Prejbeanu}}, \bibinfo {author} {\bibfnamefont {K.}~\bibnamefont {Garello}},
  \bibinfo {author} {\bibfnamefont {P.}~\bibnamefont {Gambardella}}, \bibinfo
  {author} {\bibfnamefont {P.}~\bibnamefont {Freitas}}, \bibinfo {author}
  {\bibfnamefont {R.}~\bibnamefont {Lehndorff}}, \bibinfo {author}
  {\bibfnamefont {W.}~\bibnamefont {Raberg}}, \bibinfo {author} {\bibfnamefont
  {U.}~\bibnamefont {Ebels}}, \bibinfo {author} {\bibfnamefont {S.~O.}\
  \bibnamefont {Demokritov}}, \bibinfo {author} {\bibfnamefont
  {J.}~\bibnamefont {Akerman}}, \bibinfo {author} {\bibfnamefont
  {A.}~\bibnamefont {Deac}}, \bibinfo {author} {\bibfnamefont {P.}~\bibnamefont
  {Pirro}}, \bibinfo {author} {\bibfnamefont {C.}~\bibnamefont {Adelmann}},
  \bibinfo {author} {\bibfnamefont {A.}~\bibnamefont {Anane}}, \bibinfo
  {author} {\bibfnamefont {A.~V.}\ \bibnamefont {Chumak}}, \bibinfo {author}
  {\bibfnamefont {A.}~\bibnamefont {Hirohata}}, \bibinfo {author}
  {\bibfnamefont {S.}~\bibnamefont {Mangin}}, \bibinfo {author} {\bibfnamefont
  {S.~O.}\ \bibnamefont {Valenzuela}}, \bibinfo {author} {\bibfnamefont
  {M.~C.}\ \bibnamefont {Onbaşlı}}, \bibinfo {author} {\bibfnamefont
  {M.}~\bibnamefont {d’Aquino}}, \bibinfo {author} {\bibfnamefont
  {G.}~\bibnamefont {Prenat}}, \bibinfo {author} {\bibfnamefont
  {G.}~\bibnamefont {Finocchio}}, \bibinfo {author} {\bibfnamefont
  {L.}~\bibnamefont {Lopez-Diaz}}, \bibinfo {author} {\bibfnamefont
  {R.}~\bibnamefont {Chantrell}}, \bibinfo {author} {\bibfnamefont
  {O.}~\bibnamefont {Chubykalo-Fesenko}}, \ and\ \bibinfo {author}
  {\bibfnamefont {P.}~\bibnamefont {Bortolotti}},\ }\href {\doibase
  10.1038/s41928-020-0461-5} {\bibfield  {journal} {\bibinfo  {journal} {Nature
  Electronics}\ }\textbf {\bibinfo {volume} {3}},\ \bibinfo {pages} {446}
  (\bibinfo {year} {2020})}\BibitemShut {NoStop}%
\bibitem [{\citenamefont {Moreau-Luchaire}\ \emph {et~al.}(2016)\citenamefont
  {Moreau-Luchaire}, \citenamefont {Moutafis}, \citenamefont {Reyren},
  \citenamefont {Sampaio}, \citenamefont {Vaz}, \citenamefont {Van~Horne},
  \citenamefont {Bouzehouane}, \citenamefont {Garcia}, \citenamefont
  {Deranlot}, \citenamefont {Warnicke}, \citenamefont {Wohlhuter},
  \citenamefont {George}, \citenamefont {Weigand}, \citenamefont {Raabe},
  \citenamefont {Cros},\ and\ \citenamefont {Fert}}]{MoreauLuchaire.2016}%
  \BibitemOpen
  \bibfield  {author} {\bibinfo {author} {\bibfnamefont {C.}~\bibnamefont
  {Moreau-Luchaire}}, \bibinfo {author} {\bibfnamefont {C.}~\bibnamefont
  {Moutafis}}, \bibinfo {author} {\bibfnamefont {N.}~\bibnamefont {Reyren}},
  \bibinfo {author} {\bibfnamefont {J.}~\bibnamefont {Sampaio}}, \bibinfo
  {author} {\bibfnamefont {C.~A.~F.}\ \bibnamefont {Vaz}}, \bibinfo {author}
  {\bibfnamefont {N.}~\bibnamefont {Van~Horne}}, \bibinfo {author}
  {\bibfnamefont {K.}~\bibnamefont {Bouzehouane}}, \bibinfo {author}
  {\bibfnamefont {K.}~\bibnamefont {Garcia}}, \bibinfo {author} {\bibfnamefont
  {C.}~\bibnamefont {Deranlot}}, \bibinfo {author} {\bibfnamefont
  {P.}~\bibnamefont {Warnicke}}, \bibinfo {author} {\bibfnamefont
  {P.}~\bibnamefont {Wohlhuter}}, \bibinfo {author} {\bibfnamefont {J.~M.}\
  \bibnamefont {George}}, \bibinfo {author} {\bibfnamefont {M.}~\bibnamefont
  {Weigand}}, \bibinfo {author} {\bibfnamefont {J.}~\bibnamefont {Raabe}},
  \bibinfo {author} {\bibfnamefont {V.}~\bibnamefont {Cros}}, \ and\ \bibinfo
  {author} {\bibfnamefont {A.}~\bibnamefont {Fert}},\ }\href {\doibase
  10.1038/nnano.2015.313} {\bibfield  {journal} {\bibinfo  {journal} {Nature
  Nanotechnology}\ }\textbf {\bibinfo {volume} {11}},\ \bibinfo {pages} {444}
  (\bibinfo {year} {2016})}\BibitemShut {NoStop}%
\bibitem [{\citenamefont {Woo}\ \emph {et~al.}(2016)\citenamefont {Woo},
  \citenamefont {Litzius}, \citenamefont {Kruger}, \citenamefont {Im},
  \citenamefont {Caretta}, \citenamefont {Richter}, \citenamefont {Mann},
  \citenamefont {Krone}, \citenamefont {Reeve}, \citenamefont {Weigand},
  \citenamefont {Agrawal}, \citenamefont {Lemesh}, \citenamefont {Mawass},
  \citenamefont {Fischer}, \citenamefont {Klaui},\ and\ \citenamefont
  {Beach}}]{Woo.2016}%
  \BibitemOpen
  \bibfield  {author} {\bibinfo {author} {\bibfnamefont {S.}~\bibnamefont
  {Woo}}, \bibinfo {author} {\bibfnamefont {K.}~\bibnamefont {Litzius}},
  \bibinfo {author} {\bibfnamefont {B.}~\bibnamefont {Kruger}}, \bibinfo
  {author} {\bibfnamefont {M.~Y.}\ \bibnamefont {Im}}, \bibinfo {author}
  {\bibfnamefont {L.}~\bibnamefont {Caretta}}, \bibinfo {author} {\bibfnamefont
  {K.}~\bibnamefont {Richter}}, \bibinfo {author} {\bibfnamefont
  {M.}~\bibnamefont {Mann}}, \bibinfo {author} {\bibfnamefont {A.}~\bibnamefont
  {Krone}}, \bibinfo {author} {\bibfnamefont {R.~M.}\ \bibnamefont {Reeve}},
  \bibinfo {author} {\bibfnamefont {M.}~\bibnamefont {Weigand}}, \bibinfo
  {author} {\bibfnamefont {P.}~\bibnamefont {Agrawal}}, \bibinfo {author}
  {\bibfnamefont {I.}~\bibnamefont {Lemesh}}, \bibinfo {author} {\bibfnamefont
  {M.~A.}\ \bibnamefont {Mawass}}, \bibinfo {author} {\bibfnamefont
  {P.}~\bibnamefont {Fischer}}, \bibinfo {author} {\bibfnamefont
  {M.}~\bibnamefont {Klaui}}, \ and\ \bibinfo {author} {\bibfnamefont
  {G.}~\bibnamefont {Beach}},\ }\href {\doibase 10.1038/nmat4593} {\bibfield
  {journal} {\bibinfo  {journal} {Nature Materials}\ }\textbf {\bibinfo
  {volume} {15}},\ \bibinfo {pages} {501} (\bibinfo {year} {2016})}\BibitemShut
  {NoStop}%
\bibitem [{\citenamefont {Boulle}\ \emph {et~al.}(2016)\citenamefont {Boulle},
  \citenamefont {Vogel}, \citenamefont {Yang}, \citenamefont {Pizzini},
  \citenamefont {Chaves}, \citenamefont {Locatelli}, \citenamefont {Mentes},
  \citenamefont {Sala}, \citenamefont {Buda-Prejbeanu}, \citenamefont {Klein},
  \citenamefont {Belmeguenai}, \citenamefont {Roussigne}, \citenamefont
  {Stashkevich}, \citenamefont {Cherif}, \citenamefont {Aballe}, \citenamefont
  {Foerster}, \citenamefont {Chshiev}, \citenamefont {Auffret}, \citenamefont
  {Miron},\ and\ \citenamefont {Gaudin}}]{Boulle.2016}%
  \BibitemOpen
  \bibfield  {author} {\bibinfo {author} {\bibfnamefont {O.}~\bibnamefont
  {Boulle}}, \bibinfo {author} {\bibfnamefont {J.}~\bibnamefont {Vogel}},
  \bibinfo {author} {\bibfnamefont {H.~X.}\ \bibnamefont {Yang}}, \bibinfo
  {author} {\bibfnamefont {S.}~\bibnamefont {Pizzini}}, \bibinfo {author}
  {\bibfnamefont {D.~D.}\ \bibnamefont {Chaves}}, \bibinfo {author}
  {\bibfnamefont {A.}~\bibnamefont {Locatelli}}, \bibinfo {author}
  {\bibfnamefont {T.~O.}\ \bibnamefont {Mentes}}, \bibinfo {author}
  {\bibfnamefont {A.}~\bibnamefont {Sala}}, \bibinfo {author} {\bibfnamefont
  {L.~D.}\ \bibnamefont {Buda-Prejbeanu}}, \bibinfo {author} {\bibfnamefont
  {O.}~\bibnamefont {Klein}}, \bibinfo {author} {\bibfnamefont
  {M.}~\bibnamefont {Belmeguenai}}, \bibinfo {author} {\bibfnamefont
  {Y.}~\bibnamefont {Roussigne}}, \bibinfo {author} {\bibfnamefont
  {A.}~\bibnamefont {Stashkevich}}, \bibinfo {author} {\bibfnamefont {S.~M.}\
  \bibnamefont {Cherif}}, \bibinfo {author} {\bibfnamefont {L.}~\bibnamefont
  {Aballe}}, \bibinfo {author} {\bibfnamefont {M.}~\bibnamefont {Foerster}},
  \bibinfo {author} {\bibfnamefont {M.}~\bibnamefont {Chshiev}}, \bibinfo
  {author} {\bibfnamefont {S.}~\bibnamefont {Auffret}}, \bibinfo {author}
  {\bibfnamefont {I.~M.}\ \bibnamefont {Miron}}, \ and\ \bibinfo {author}
  {\bibfnamefont {G.}~\bibnamefont {Gaudin}},\ }\href {\doibase
  10.1038/nnano.2015.315} {\bibfield  {journal} {\bibinfo  {journal} {Nature
  Nanotechnology}\ }\textbf {\bibinfo {volume} {11}},\ \bibinfo {pages} {449}
  (\bibinfo {year} {2016})}\BibitemShut {NoStop}%
\bibitem [{\citenamefont {Soumyanarayanan}\ \emph {et~al.}(2017)\citenamefont
  {Soumyanarayanan}, \citenamefont {Raju}, \citenamefont {Oyarce},
  \citenamefont {Tan}, \citenamefont {Im}, \citenamefont {Petrovic},
  \citenamefont {Ho}, \citenamefont {Khoo}, \citenamefont {Tran}, \citenamefont
  {Gan}, \citenamefont {Ernult},\ and\ \citenamefont
  {Panagopoulos}}]{Soumyanarayanan.2017}%
  \BibitemOpen
  \bibfield  {author} {\bibinfo {author} {\bibfnamefont {A.}~\bibnamefont
  {Soumyanarayanan}}, \bibinfo {author} {\bibfnamefont {M.}~\bibnamefont
  {Raju}}, \bibinfo {author} {\bibfnamefont {A.~L.~G.}\ \bibnamefont {Oyarce}},
  \bibinfo {author} {\bibfnamefont {A.~K.~C.}\ \bibnamefont {Tan}}, \bibinfo
  {author} {\bibfnamefont {M.~Y.}\ \bibnamefont {Im}}, \bibinfo {author}
  {\bibfnamefont {A.~P.}\ \bibnamefont {Petrovic}}, \bibinfo {author}
  {\bibfnamefont {P.}~\bibnamefont {Ho}}, \bibinfo {author} {\bibfnamefont
  {K.~H.}\ \bibnamefont {Khoo}}, \bibinfo {author} {\bibfnamefont
  {M.}~\bibnamefont {Tran}}, \bibinfo {author} {\bibfnamefont {C.~K.}\
  \bibnamefont {Gan}}, \bibinfo {author} {\bibfnamefont {F.}~\bibnamefont
  {Ernult}}, \ and\ \bibinfo {author} {\bibfnamefont {C.}~\bibnamefont
  {Panagopoulos}},\ }\href {\doibase 10.1038/nmat4934} {\bibfield  {journal}
  {\bibinfo  {journal} {Nature Materials}\ }\textbf {\bibinfo {volume} {16}},\
  \bibinfo {pages} {898} (\bibinfo {year} {2017})}\BibitemShut {NoStop}%
\bibitem [{\citenamefont {Nagaosa}\ and\ \citenamefont
  {Tokura}(2013)}]{Nagaosa.2013}%
  \BibitemOpen
  \bibfield  {author} {\bibinfo {author} {\bibfnamefont {N.}~\bibnamefont
  {Nagaosa}}\ and\ \bibinfo {author} {\bibfnamefont {Y.}~\bibnamefont
  {Tokura}},\ }\href {\doibase 10.1038/nnano.2013.243} {\bibfield  {journal}
  {\bibinfo  {journal} {Nature Nanotechnology}\ }\textbf {\bibinfo {volume}
  {8}},\ \bibinfo {pages} {899} (\bibinfo {year} {2013})}\BibitemShut {NoStop}%
\bibitem [{\citenamefont {Jiang}\ \emph {et~al.}(2015)\citenamefont {Jiang},
  \citenamefont {Upadhyaya}, \citenamefont {Zhang}, \citenamefont {Yu},
  \citenamefont {Jungfleisch}, \citenamefont {Fradin}, \citenamefont {Pearson},
  \citenamefont {Tserkovnyak}, \citenamefont {Wang}, \citenamefont {Heinonen},
  \citenamefont {te~Velthuis},\ and\ \citenamefont {Hoffmann}}]{Jiang.2015}%
  \BibitemOpen
  \bibfield  {author} {\bibinfo {author} {\bibfnamefont {W.~J.}\ \bibnamefont
  {Jiang}}, \bibinfo {author} {\bibfnamefont {P.}~\bibnamefont {Upadhyaya}},
  \bibinfo {author} {\bibfnamefont {W.}~\bibnamefont {Zhang}}, \bibinfo
  {author} {\bibfnamefont {G.~Q.}\ \bibnamefont {Yu}}, \bibinfo {author}
  {\bibfnamefont {M.~B.}\ \bibnamefont {Jungfleisch}}, \bibinfo {author}
  {\bibfnamefont {F.~Y.}\ \bibnamefont {Fradin}}, \bibinfo {author}
  {\bibfnamefont {J.~E.}\ \bibnamefont {Pearson}}, \bibinfo {author}
  {\bibfnamefont {Y.}~\bibnamefont {Tserkovnyak}}, \bibinfo {author}
  {\bibfnamefont {K.~L.}\ \bibnamefont {Wang}}, \bibinfo {author}
  {\bibfnamefont {O.}~\bibnamefont {Heinonen}}, \bibinfo {author}
  {\bibfnamefont {S.~G.~E.}\ \bibnamefont {te~Velthuis}}, \ and\ \bibinfo
  {author} {\bibfnamefont {A.}~\bibnamefont {Hoffmann}},\ }\href {\doibase
  10.1126/science.aaa1442} {\bibfield  {journal} {\bibinfo  {journal}
  {Science}\ }\textbf {\bibinfo {volume} {349}},\ \bibinfo {pages} {283}
  (\bibinfo {year} {2015})}\BibitemShut {NoStop}%
\bibitem [{\citenamefont {Zázvorka}\ \emph {et~al.}(2019)\citenamefont
  {Zázvorka}, \citenamefont {Jakobs}, \citenamefont {Heinze}, \citenamefont
  {Keil}, \citenamefont {Kromin}, \citenamefont {Jaiswal}, \citenamefont
  {Litzius}, \citenamefont {Jakob}, \citenamefont {Virnau}, \citenamefont
  {Pinna}, \citenamefont {Everschor-Sitte}, \citenamefont {Rózsa},
  \citenamefont {Donges}, \citenamefont {Nowak},\ and\ \citenamefont
  {Kläui}}]{Zazvorka.2019}%
  \BibitemOpen
  \bibfield  {author} {\bibinfo {author} {\bibfnamefont {J.}~\bibnamefont
  {Zázvorka}}, \bibinfo {author} {\bibfnamefont {F.}~\bibnamefont {Jakobs}},
  \bibinfo {author} {\bibfnamefont {D.}~\bibnamefont {Heinze}}, \bibinfo
  {author} {\bibfnamefont {N.}~\bibnamefont {Keil}}, \bibinfo {author}
  {\bibfnamefont {S.}~\bibnamefont {Kromin}}, \bibinfo {author} {\bibfnamefont
  {S.}~\bibnamefont {Jaiswal}}, \bibinfo {author} {\bibfnamefont
  {K.}~\bibnamefont {Litzius}}, \bibinfo {author} {\bibfnamefont
  {G.}~\bibnamefont {Jakob}}, \bibinfo {author} {\bibfnamefont
  {P.}~\bibnamefont {Virnau}}, \bibinfo {author} {\bibfnamefont
  {D.}~\bibnamefont {Pinna}}, \bibinfo {author} {\bibfnamefont
  {K.}~\bibnamefont {Everschor-Sitte}}, \bibinfo {author} {\bibfnamefont
  {L.}~\bibnamefont {Rózsa}}, \bibinfo {author} {\bibfnamefont
  {A.}~\bibnamefont {Donges}}, \bibinfo {author} {\bibfnamefont
  {U.}~\bibnamefont {Nowak}}, \ and\ \bibinfo {author} {\bibfnamefont
  {M.}~\bibnamefont {Kläui}},\ }\href {\doibase 10.1038/s41565-019-0436-8}
  {\bibfield  {journal} {\bibinfo  {journal} {Nature Nanotechnology}\ }\textbf
  {\bibinfo {volume} {14}},\ \bibinfo {pages} {658} (\bibinfo {year}
  {2019})}\BibitemShut {NoStop}%
\bibitem [{\citenamefont {Grollier}\ \emph {et~al.}(2020)\citenamefont
  {Grollier}, \citenamefont {Querlioz}, \citenamefont {Camsari}, \citenamefont
  {Everschor-Sitte}, \citenamefont {Fukami},\ and\ \citenamefont
  {Stiles}}]{Grollier.2020}%
  \BibitemOpen
  \bibfield  {author} {\bibinfo {author} {\bibfnamefont {J.}~\bibnamefont
  {Grollier}}, \bibinfo {author} {\bibfnamefont {D.}~\bibnamefont {Querlioz}},
  \bibinfo {author} {\bibfnamefont {K.~Y.}\ \bibnamefont {Camsari}}, \bibinfo
  {author} {\bibfnamefont {K.}~\bibnamefont {Everschor-Sitte}}, \bibinfo
  {author} {\bibfnamefont {S.}~\bibnamefont {Fukami}}, \ and\ \bibinfo {author}
  {\bibfnamefont {M.~D.}\ \bibnamefont {Stiles}},\ }\href {\doibase
  10.1038/s41928-019-0360-9} {\bibfield  {journal} {\bibinfo  {journal} {Nature
  Electronics}\ }\textbf {\bibinfo {volume} {3}},\ \bibinfo {pages} {360}
  (\bibinfo {year} {2020})}\BibitemShut {NoStop}%
\bibitem [{\citenamefont {Fert}\ \emph {et~al.}(2017)\citenamefont {Fert},
  \citenamefont {Reyren},\ and\ \citenamefont {Cros}}]{Fert.2017}%
  \BibitemOpen
  \bibfield  {author} {\bibinfo {author} {\bibfnamefont {A.}~\bibnamefont
  {Fert}}, \bibinfo {author} {\bibfnamefont {N.}~\bibnamefont {Reyren}}, \ and\
  \bibinfo {author} {\bibfnamefont {V.}~\bibnamefont {Cros}},\ }\href {\doibase
  10.1038/natrevmats.2017.31} {\bibfield  {journal} {\bibinfo  {journal}
  {Nature Reviews Materials}\ }\textbf {\bibinfo {volume} {2}},\ \bibinfo
  {pages} {17031} (\bibinfo {year} {2017})}\BibitemShut {NoStop}%
\bibitem [{\citenamefont {Romming}\ \emph {et~al.}(2013)\citenamefont
  {Romming}, \citenamefont {Hanneken}, \citenamefont {Menzel}, \citenamefont
  {Bickel}, \citenamefont {Wolter}, \citenamefont {von Bergmann}, \citenamefont
  {Kubetzka},\ and\ \citenamefont {Wiesendanger}}]{Romming.2013}%
  \BibitemOpen
  \bibfield  {author} {\bibinfo {author} {\bibfnamefont {N.}~\bibnamefont
  {Romming}}, \bibinfo {author} {\bibfnamefont {C.}~\bibnamefont {Hanneken}},
  \bibinfo {author} {\bibfnamefont {M.}~\bibnamefont {Menzel}}, \bibinfo
  {author} {\bibfnamefont {J.~E.}\ \bibnamefont {Bickel}}, \bibinfo {author}
  {\bibfnamefont {B.}~\bibnamefont {Wolter}}, \bibinfo {author} {\bibfnamefont
  {K.}~\bibnamefont {von Bergmann}}, \bibinfo {author} {\bibfnamefont
  {A.}~\bibnamefont {Kubetzka}}, \ and\ \bibinfo {author} {\bibfnamefont
  {R.}~\bibnamefont {Wiesendanger}},\ }\href {\doibase 10.1126/science.1240573}
  {\bibfield  {journal} {\bibinfo  {journal} {Science}\ }\textbf {\bibinfo
  {volume} {341}},\ \bibinfo {pages} {636} (\bibinfo {year}
  {2013})}\BibitemShut {NoStop}%
\bibitem [{\citenamefont {Roy}\ \emph {et~al.}(2019)\citenamefont {Roy},
  \citenamefont {Jaiswal},\ and\ \citenamefont {Panda}}]{Roy.2019}%
  \BibitemOpen
  \bibfield  {author} {\bibinfo {author} {\bibfnamefont {K.}~\bibnamefont
  {Roy}}, \bibinfo {author} {\bibfnamefont {A.}~\bibnamefont {Jaiswal}}, \ and\
  \bibinfo {author} {\bibfnamefont {P.}~\bibnamefont {Panda}},\ }\href
  {\doibase 10.1038/s41586-019-1677-2} {\bibfield  {journal} {\bibinfo
  {journal} {Nature}\ }\textbf {\bibinfo {volume} {575}},\ \bibinfo {pages}
  {607} (\bibinfo {year} {2019})}\BibitemShut {NoStop}%
\bibitem [{\citenamefont {Guang}\ \emph {et~al.}(2022)\citenamefont {Guang},
  \citenamefont {Zhang}, \citenamefont {Zhang}, \citenamefont {Wang},
  \citenamefont {Zhao}, \citenamefont {Tomasello}, \citenamefont {Zhang},
  \citenamefont {He}, \citenamefont {Li}, \citenamefont {Liu}, \citenamefont
  {Feng}, \citenamefont {Wei}, \citenamefont {Carpentieri}, \citenamefont
  {Hou}, \citenamefont {Liu}, \citenamefont {Peng}, \citenamefont {Zeng},
  \citenamefont {Finocchio}, \citenamefont {Zhang}, \citenamefont {Coey},
  \citenamefont {Han},\ and\ \citenamefont {Yu}}]{Guang.2022}%
  \BibitemOpen
  \bibfield  {author} {\bibinfo {author} {\bibfnamefont {Y.}~\bibnamefont
  {Guang}}, \bibinfo {author} {\bibfnamefont {L.}~\bibnamefont {Zhang}},
  \bibinfo {author} {\bibfnamefont {J.}~\bibnamefont {Zhang}}, \bibinfo
  {author} {\bibfnamefont {Y.}~\bibnamefont {Wang}}, \bibinfo {author}
  {\bibfnamefont {Y.}~\bibnamefont {Zhao}}, \bibinfo {author} {\bibfnamefont
  {R.}~\bibnamefont {Tomasello}}, \bibinfo {author} {\bibfnamefont
  {S.}~\bibnamefont {Zhang}}, \bibinfo {author} {\bibfnamefont
  {B.}~\bibnamefont {He}}, \bibinfo {author} {\bibfnamefont {J.}~\bibnamefont
  {Li}}, \bibinfo {author} {\bibfnamefont {Y.}~\bibnamefont {Liu}}, \bibinfo
  {author} {\bibfnamefont {J.}~\bibnamefont {Feng}}, \bibinfo {author}
  {\bibfnamefont {H.}~\bibnamefont {Wei}}, \bibinfo {author} {\bibfnamefont
  {M.}~\bibnamefont {Carpentieri}}, \bibinfo {author} {\bibfnamefont
  {Z.}~\bibnamefont {Hou}}, \bibinfo {author} {\bibfnamefont {J.}~\bibnamefont
  {Liu}}, \bibinfo {author} {\bibfnamefont {Y.}~\bibnamefont {Peng}}, \bibinfo
  {author} {\bibfnamefont {Z.}~\bibnamefont {Zeng}}, \bibinfo {author}
  {\bibfnamefont {G.}~\bibnamefont {Finocchio}}, \bibinfo {author}
  {\bibfnamefont {X.}~\bibnamefont {Zhang}}, \bibinfo {author} {\bibfnamefont
  {J.~M.~D.}\ \bibnamefont {Coey}}, \bibinfo {author} {\bibfnamefont
  {X.}~\bibnamefont {Han}}, \ and\ \bibinfo {author} {\bibfnamefont
  {G.}~\bibnamefont {Yu}},\ }\href {\doibase
  https://doi.org/10.1002/aelm.202200570} {\bibfield  {journal} {\bibinfo
  {journal} {Advanced Electronic Materials}\ }\textbf {\bibinfo {volume} {9}},\
  \bibinfo {pages} {2200570} (\bibinfo {year} {2022})}\BibitemShut {NoStop}%
\bibitem [{\citenamefont {Li}\ \emph {et~al.}(2022)\citenamefont {Li},
  \citenamefont {Du}, \citenamefont {Wang}, \citenamefont {Wang}, \citenamefont
  {Zhang}, \citenamefont {Cheng}, \citenamefont {Cai}, \citenamefont {Lu},
  \citenamefont {Cao}, \citenamefont {Pan}, \citenamefont {Lei}, \citenamefont
  {Kang}, \citenamefont {Liu}, \citenamefont {Fert}, \citenamefont {Hou},\ and\
  \citenamefont {Zhao}}]{Li.2022}%
  \BibitemOpen
  \bibfield  {author} {\bibinfo {author} {\bibfnamefont {S.}~\bibnamefont
  {Li}}, \bibinfo {author} {\bibfnamefont {A.}~\bibnamefont {Du}}, \bibinfo
  {author} {\bibfnamefont {Y.}~\bibnamefont {Wang}}, \bibinfo {author}
  {\bibfnamefont {X.}~\bibnamefont {Wang}}, \bibinfo {author} {\bibfnamefont
  {X.}~\bibnamefont {Zhang}}, \bibinfo {author} {\bibfnamefont
  {H.}~\bibnamefont {Cheng}}, \bibinfo {author} {\bibfnamefont
  {W.}~\bibnamefont {Cai}}, \bibinfo {author} {\bibfnamefont {S.}~\bibnamefont
  {Lu}}, \bibinfo {author} {\bibfnamefont {K.}~\bibnamefont {Cao}}, \bibinfo
  {author} {\bibfnamefont {B.}~\bibnamefont {Pan}}, \bibinfo {author}
  {\bibfnamefont {N.}~\bibnamefont {Lei}}, \bibinfo {author} {\bibfnamefont
  {W.}~\bibnamefont {Kang}}, \bibinfo {author} {\bibfnamefont {J.}~\bibnamefont
  {Liu}}, \bibinfo {author} {\bibfnamefont {A.}~\bibnamefont {Fert}}, \bibinfo
  {author} {\bibfnamefont {Z.}~\bibnamefont {Hou}}, \ and\ \bibinfo {author}
  {\bibfnamefont {W.}~\bibnamefont {Zhao}},\ }\href {\doibase
  https://doi.org/10.1016/j.scib.2022.01.016} {\bibfield  {journal} {\bibinfo
  {journal} {Science Bulletin}\ }\textbf {\bibinfo {volume} {67}},\ \bibinfo
  {pages} {691} (\bibinfo {year} {2022})}\BibitemShut {NoStop}%
\bibitem [{\citenamefont {Kasai}\ \emph {et~al.}(2019)\citenamefont {Kasai},
  \citenamefont {Sugimoto}, \citenamefont {Nakatani}, \citenamefont
  {Ishikawa},\ and\ \citenamefont {Takahashi}}]{Kasai.2019}%
  \BibitemOpen
  \bibfield  {author} {\bibinfo {author} {\bibfnamefont {S.}~\bibnamefont
  {Kasai}}, \bibinfo {author} {\bibfnamefont {S.}~\bibnamefont {Sugimoto}},
  \bibinfo {author} {\bibfnamefont {Y.}~\bibnamefont {Nakatani}}, \bibinfo
  {author} {\bibfnamefont {R.}~\bibnamefont {Ishikawa}}, \ and\ \bibinfo
  {author} {\bibfnamefont {Y.~K.}\ \bibnamefont {Takahashi}},\ }\href {\doibase
  10.7567/1882-0786/ab2baa} {\bibfield  {journal} {\bibinfo  {journal} {Applied
  Physics Express}\ }\textbf {\bibinfo {volume} {12}},\ \bibinfo {pages}
  {083001} (\bibinfo {year} {2019})}\BibitemShut {NoStop}%
\bibitem [{\citenamefont {Penthorn}\ \emph {et~al.}(2019)\citenamefont
  {Penthorn}, \citenamefont {Hao}, \citenamefont {Wang}, \citenamefont {Huai},\
  and\ \citenamefont {Jiang}}]{Penthorn.2019}%
  \BibitemOpen
  \bibfield  {author} {\bibinfo {author} {\bibfnamefont {N.~E.}\ \bibnamefont
  {Penthorn}}, \bibinfo {author} {\bibfnamefont {X.}~\bibnamefont {Hao}},
  \bibinfo {author} {\bibfnamefont {Z.}~\bibnamefont {Wang}}, \bibinfo {author}
  {\bibfnamefont {Y.}~\bibnamefont {Huai}}, \ and\ \bibinfo {author}
  {\bibfnamefont {H.~W.}\ \bibnamefont {Jiang}},\ }\href {\doibase
  10.1103/PhysRevLett.122.257201} {\bibfield  {journal} {\bibinfo  {journal}
  {Physical Review Letters}\ }\textbf {\bibinfo {volume} {122}},\ \bibinfo
  {pages} {257201} (\bibinfo {year} {2019})}\BibitemShut {NoStop}%
\bibitem [{\citenamefont {Kim}\ \emph {et~al.}(2017)\citenamefont {Kim},
  \citenamefont {Park},\ and\ \citenamefont {Park}}]{Kim.2017}%
  \BibitemOpen
  \bibfield  {author} {\bibinfo {author} {\bibfnamefont {D.-H.}\ \bibnamefont
  {Kim}}, \bibinfo {author} {\bibfnamefont {K.-W.}\ \bibnamefont {Park}}, \
  and\ \bibinfo {author} {\bibfnamefont {B.-G.}\ \bibnamefont {Park}},\ }\href
  {\doibase https://doi.org/10.1016/j.cap.2017.04.003} {\bibfield  {journal}
  {\bibinfo  {journal} {Current Applied Physics}\ }\textbf {\bibinfo {volume}
  {17}},\ \bibinfo {pages} {962} (\bibinfo {year} {2017})}\BibitemShut
  {NoStop}%
\bibitem [{\citenamefont {Chen}\ \emph
  {et~al.}(2022{\natexlab{a}})\citenamefont {Chen}, \citenamefont {Lin},
  \citenamefont {Kong}, \citenamefont {Tan}, \citenamefont {Tan}, \citenamefont
  {Je}, \citenamefont {Tan}, \citenamefont {Khoo}, \citenamefont {Im},\ and\
  \citenamefont {Soumyanarayanan}}]{Chen.2022}%
  \BibitemOpen
  \bibfield  {author} {\bibinfo {author} {\bibfnamefont {X.}~\bibnamefont
  {Chen}}, \bibinfo {author} {\bibfnamefont {M.}~\bibnamefont {Lin}}, \bibinfo
  {author} {\bibfnamefont {J.~F.}\ \bibnamefont {Kong}}, \bibinfo {author}
  {\bibfnamefont {H.~R.}\ \bibnamefont {Tan}}, \bibinfo {author} {\bibfnamefont
  {A.~K.~C.}\ \bibnamefont {Tan}}, \bibinfo {author} {\bibfnamefont {S.-G.}\
  \bibnamefont {Je}}, \bibinfo {author} {\bibfnamefont {H.~K.}\ \bibnamefont
  {Tan}}, \bibinfo {author} {\bibfnamefont {K.~H.}\ \bibnamefont {Khoo}},
  \bibinfo {author} {\bibfnamefont {M.-Y.}\ \bibnamefont {Im}}, \ and\ \bibinfo
  {author} {\bibfnamefont {A.}~\bibnamefont {Soumyanarayanan}},\ }\href
  {\doibase https://doi.org/10.1002/advs.202103978} {\bibfield  {journal}
  {\bibinfo  {journal} {Advanced Science}\ }\textbf {\bibinfo {volume} {9}},\
  \bibinfo {pages} {2103978} (\bibinfo {year}
  {2022}{\natexlab{a}})}\BibitemShut {NoStop}%
\bibitem [{\citenamefont {Han}\ \emph {et~al.}(2015)\citenamefont {Han},
  \citenamefont {Tran}, \citenamefont {Sim}, \citenamefont {Wang},
  \citenamefont {Eason}, \citenamefont {Lim},\ and\ \citenamefont
  {Huang}}]{Han.2015}%
  \BibitemOpen
  \bibfield  {author} {\bibinfo {author} {\bibfnamefont {G.}~\bibnamefont
  {Han}}, \bibinfo {author} {\bibfnamefont {M.}~\bibnamefont {Tran}}, \bibinfo
  {author} {\bibfnamefont {C.~H.}\ \bibnamefont {Sim}}, \bibinfo {author}
  {\bibfnamefont {J.~C.}\ \bibnamefont {Wang}}, \bibinfo {author}
  {\bibfnamefont {K.}~\bibnamefont {Eason}}, \bibinfo {author} {\bibfnamefont
  {S.~T.}\ \bibnamefont {Lim}}, \ and\ \bibinfo {author} {\bibfnamefont
  {A.}~\bibnamefont {Huang}},\ }\href {\doibase 10.1063/1.4913942} {\bibfield
  {journal} {\bibinfo  {journal} {Journal of Applied Physics}\ }\textbf
  {\bibinfo {volume} {117}},\ \bibinfo {pages} {17B515} (\bibinfo {year}
  {2015})}\BibitemShut {NoStop}%
\bibitem [{\citenamefont {Lim}\ \emph {et~al.}(2015)\citenamefont {Lim},
  \citenamefont {Tran}, \citenamefont {Chenchen}, \citenamefont {Ying},\ and\
  \citenamefont {Han}}]{Lim.2015}%
  \BibitemOpen
  \bibfield  {author} {\bibinfo {author} {\bibfnamefont {S.~T.}\ \bibnamefont
  {Lim}}, \bibinfo {author} {\bibfnamefont {M.}~\bibnamefont {Tran}}, \bibinfo
  {author} {\bibfnamefont {J.~W.}\ \bibnamefont {Chenchen}}, \bibinfo {author}
  {\bibfnamefont {J.~F.}\ \bibnamefont {Ying}}, \ and\ \bibinfo {author}
  {\bibfnamefont {G.}~\bibnamefont {Han}},\ }\href {\doibase 10.1063/1.4916295}
  {\bibfield  {journal} {\bibinfo  {journal} {Journal of Applied Physics}\
  }\textbf {\bibinfo {volume} {117}},\ \bibinfo {pages} {17A731} (\bibinfo
  {year} {2015})}\BibitemShut {NoStop}%
\bibitem [{\citenamefont {Ho}\ \emph {et~al.}(2019)\citenamefont {Ho},
  \citenamefont {Tan}, \citenamefont {Goolaup}, \citenamefont {Oyarce},
  \citenamefont {Raju}, \citenamefont {Huang}, \citenamefont
  {Soumyanarayanan},\ and\ \citenamefont {Panagopoulos}}]{Ho.2019}%
  \BibitemOpen
  \bibfield  {author} {\bibinfo {author} {\bibfnamefont {P.}~\bibnamefont
  {Ho}}, \bibinfo {author} {\bibfnamefont {A.~K.~C.}\ \bibnamefont {Tan}},
  \bibinfo {author} {\bibfnamefont {S.}~\bibnamefont {Goolaup}}, \bibinfo
  {author} {\bibfnamefont {A.~L.~G.}\ \bibnamefont {Oyarce}}, \bibinfo {author}
  {\bibfnamefont {M.}~\bibnamefont {Raju}}, \bibinfo {author} {\bibfnamefont
  {L.~S.}\ \bibnamefont {Huang}}, \bibinfo {author} {\bibfnamefont
  {A.}~\bibnamefont {Soumyanarayanan}}, \ and\ \bibinfo {author} {\bibfnamefont
  {C.}~\bibnamefont {Panagopoulos}},\ }\href {\doibase
  10.1103/PhysRevApplied.11.024064} {\bibfield  {journal} {\bibinfo  {journal}
  {Physical Review Applied}\ }\textbf {\bibinfo {volume} {11}},\ \bibinfo
  {pages} {024064} (\bibinfo {year} {2019})}\BibitemShut {NoStop}%
\bibitem [{\citenamefont {Chen}\ \emph
  {et~al.}(2022{\natexlab{b}})\citenamefont {Chen}, \citenamefont {Bouckaert},\
  and\ \citenamefont {Majetich}}]{Chen.2022b}%
  \BibitemOpen
  \bibfield  {author} {\bibinfo {author} {\bibfnamefont {H.}~\bibnamefont
  {Chen}}, \bibinfo {author} {\bibfnamefont {W.}~\bibnamefont {Bouckaert}}, \
  and\ \bibinfo {author} {\bibfnamefont {S.~A.}\ \bibnamefont {Majetich}},\
  }\href {\doibase 10.1016/j.jmmm.2021.168552} {\bibfield  {journal} {\bibinfo
  {journal} {Journal of Magnetism and Magnetic Materials}\ }\textbf {\bibinfo
  {volume} {541}},\ \bibinfo {pages} {168552} (\bibinfo {year}
  {2022}{\natexlab{b}})}\BibitemShut {NoStop}%
\bibitem [{\citenamefont {Zeissler}\ \emph {et~al.}(2017)\citenamefont
  {Zeissler}, \citenamefont {Mruczkiewicz}, \citenamefont {Finizio},
  \citenamefont {Raabe}, \citenamefont {Shepley}, \citenamefont {Sadovnikov},
  \citenamefont {Nikitov}, \citenamefont {Fallon}, \citenamefont {McFadzean},
  \citenamefont {McVitie}, \citenamefont {Moore}, \citenamefont {Burnell},\
  and\ \citenamefont {Marrows}}]{Zeissler.2017}%
  \BibitemOpen
  \bibfield  {author} {\bibinfo {author} {\bibfnamefont {K.}~\bibnamefont
  {Zeissler}}, \bibinfo {author} {\bibfnamefont {M.}~\bibnamefont
  {Mruczkiewicz}}, \bibinfo {author} {\bibfnamefont {S.}~\bibnamefont
  {Finizio}}, \bibinfo {author} {\bibfnamefont {J.}~\bibnamefont {Raabe}},
  \bibinfo {author} {\bibfnamefont {P.~M.}\ \bibnamefont {Shepley}}, \bibinfo
  {author} {\bibfnamefont {A.~V.}\ \bibnamefont {Sadovnikov}}, \bibinfo
  {author} {\bibfnamefont {S.~A.}\ \bibnamefont {Nikitov}}, \bibinfo {author}
  {\bibfnamefont {K.}~\bibnamefont {Fallon}}, \bibinfo {author} {\bibfnamefont
  {S.}~\bibnamefont {McFadzean}}, \bibinfo {author} {\bibfnamefont
  {S.}~\bibnamefont {McVitie}}, \bibinfo {author} {\bibfnamefont {T.~A.}\
  \bibnamefont {Moore}}, \bibinfo {author} {\bibfnamefont {G.}~\bibnamefont
  {Burnell}}, \ and\ \bibinfo {author} {\bibfnamefont {C.~H.}\ \bibnamefont
  {Marrows}},\ }\href {\doibase 10.1038/s41598-017-15262-3} {\bibfield
  {journal} {\bibinfo  {journal} {Scientific Reports}\ }\textbf {\bibinfo
  {volume} {7}},\ \bibinfo {pages} {15125} (\bibinfo {year}
  {2017})}\BibitemShut {NoStop}%
\bibitem [{\citenamefont {Zhang}\ \emph {et~al.}(2018)\citenamefont {Zhang},
  \citenamefont {Cai}, \citenamefont {Zhang}, \citenamefont {Wang},
  \citenamefont {Li}, \citenamefont {Zhang}, \citenamefont {Cao}, \citenamefont
  {Lei}, \citenamefont {Kang}, \citenamefont {Zhang}, \citenamefont {Yu},
  \citenamefont {Zhou},\ and\ \citenamefont {Zhao}}]{Zhang.2018}%
  \BibitemOpen
  \bibfield  {author} {\bibinfo {author} {\bibfnamefont {X.}~\bibnamefont
  {Zhang}}, \bibinfo {author} {\bibfnamefont {W.}~\bibnamefont {Cai}}, \bibinfo
  {author} {\bibfnamefont {X.}~\bibnamefont {Zhang}}, \bibinfo {author}
  {\bibfnamefont {Z.}~\bibnamefont {Wang}}, \bibinfo {author} {\bibfnamefont
  {Z.}~\bibnamefont {Li}}, \bibinfo {author} {\bibfnamefont {Y.}~\bibnamefont
  {Zhang}}, \bibinfo {author} {\bibfnamefont {K.}~\bibnamefont {Cao}}, \bibinfo
  {author} {\bibfnamefont {N.}~\bibnamefont {Lei}}, \bibinfo {author}
  {\bibfnamefont {W.}~\bibnamefont {Kang}}, \bibinfo {author} {\bibfnamefont
  {Y.}~\bibnamefont {Zhang}}, \bibinfo {author} {\bibfnamefont
  {H.}~\bibnamefont {Yu}}, \bibinfo {author} {\bibfnamefont {Y.}~\bibnamefont
  {Zhou}}, \ and\ \bibinfo {author} {\bibfnamefont {W.}~\bibnamefont {Zhao}},\
  }\href {\doibase 10.1021/acsami.8b03812} {\bibfield  {journal} {\bibinfo
  {journal} {ACS Applied Materials \& Interfaces}\ }\textbf {\bibinfo {volume}
  {10}},\ \bibinfo {pages} {16887} (\bibinfo {year} {2018})}\BibitemShut
  {NoStop}%
\bibitem [{\citenamefont {Davies}\ \emph {et~al.}(2004)\citenamefont {Davies},
  \citenamefont {Hellwig}, \citenamefont {Fullerton}, \citenamefont {Denbeaux},
  \citenamefont {Kortright},\ and\ \citenamefont {Liu}}]{Davies.2004}%
  \BibitemOpen
  \bibfield  {author} {\bibinfo {author} {\bibfnamefont {J.~E.}\ \bibnamefont
  {Davies}}, \bibinfo {author} {\bibfnamefont {O.}~\bibnamefont {Hellwig}},
  \bibinfo {author} {\bibfnamefont {E.~E.}\ \bibnamefont {Fullerton}}, \bibinfo
  {author} {\bibfnamefont {G.}~\bibnamefont {Denbeaux}}, \bibinfo {author}
  {\bibfnamefont {J.~B.}\ \bibnamefont {Kortright}}, \ and\ \bibinfo {author}
  {\bibfnamefont {K.}~\bibnamefont {Liu}},\ }\href {\doibase
  10.1103/PhysRevB.70.224434} {\bibfield  {journal} {\bibinfo  {journal}
  {Physical Review B}\ }\textbf {\bibinfo {volume} {70}},\ \bibinfo {pages}
  {224434} (\bibinfo {year} {2004})}\BibitemShut {NoStop}%
\bibitem [{\citenamefont {Tan}\ \emph {et~al.}(2020)\citenamefont {Tan},
  \citenamefont {Lourembam}, \citenamefont {Chen}, \citenamefont {Ho},
  \citenamefont {Tan},\ and\ \citenamefont {Soumyanarayanan}}]{Tan.2020}%
  \BibitemOpen
  \bibfield  {author} {\bibinfo {author} {\bibfnamefont {A.~K.~C.}\
  \bibnamefont {Tan}}, \bibinfo {author} {\bibfnamefont {J.}~\bibnamefont
  {Lourembam}}, \bibinfo {author} {\bibfnamefont {X.~Y.}\ \bibnamefont {Chen}},
  \bibinfo {author} {\bibfnamefont {P.}~\bibnamefont {Ho}}, \bibinfo {author}
  {\bibfnamefont {H.~K.}\ \bibnamefont {Tan}}, \ and\ \bibinfo {author}
  {\bibfnamefont {A.}~\bibnamefont {Soumyanarayanan}},\ }\href {\doibase
  10.1103/PhysRevMaterials.4.114419} {\bibfield  {journal} {\bibinfo  {journal}
  {Physical Review Materials}\ }\textbf {\bibinfo {volume} {4}},\ \bibinfo
  {pages} {7} (\bibinfo {year} {2020})}\BibitemShut {NoStop}%
\bibitem [{\citenamefont {Pomeroy}\ \emph {et~al.}(2009)\citenamefont
  {Pomeroy}, \citenamefont {White}, \citenamefont {Grube}, \citenamefont
  {Read},\ and\ \citenamefont {Davies}}]{Pomeroy.2009}%
  \BibitemOpen
  \bibfield  {author} {\bibinfo {author} {\bibfnamefont {J.~M.}\ \bibnamefont
  {Pomeroy}}, \bibinfo {author} {\bibfnamefont {T.~C.}\ \bibnamefont {White}},
  \bibinfo {author} {\bibfnamefont {H.}~\bibnamefont {Grube}}, \bibinfo
  {author} {\bibfnamefont {J.~C.}\ \bibnamefont {Read}}, \ and\ \bibinfo
  {author} {\bibfnamefont {J.~E.}\ \bibnamefont {Davies}},\ }\href {\doibase
  10.1063/1.3175723} {\bibfield  {journal} {\bibinfo  {journal} {Applied
  Physics Letters}\ }\textbf {\bibinfo {volume} {95}},\ \bibinfo {pages}
  {022514} (\bibinfo {year} {2009})}\BibitemShut {NoStop}%
\bibitem [{\citenamefont {Watanabe}\ \emph {et~al.}(2018)\citenamefont
  {Watanabe}, \citenamefont {Jinnai}, \citenamefont {Fukami}, \citenamefont
  {Sato},\ and\ \citenamefont {Ohno}}]{Watanabe.2018}%
  \BibitemOpen
  \bibfield  {author} {\bibinfo {author} {\bibfnamefont {K.}~\bibnamefont
  {Watanabe}}, \bibinfo {author} {\bibfnamefont {B.}~\bibnamefont {Jinnai}},
  \bibinfo {author} {\bibfnamefont {S.}~\bibnamefont {Fukami}}, \bibinfo
  {author} {\bibfnamefont {H.}~\bibnamefont {Sato}}, \ and\ \bibinfo {author}
  {\bibfnamefont {H.}~\bibnamefont {Ohno}},\ }\href {\doibase
  10.1038/s41467-018-03003-7} {\bibfield  {journal} {\bibinfo  {journal}
  {Nature Communications}\ }\textbf {\bibinfo {volume} {9}},\ \bibinfo {pages}
  {663} (\bibinfo {year} {2018})}\BibitemShut {NoStop}%
\bibitem [{\citenamefont {Liu}\ \emph {et~al.}(2012)\citenamefont {Liu},
  \citenamefont {Pai}, \citenamefont {Li}, \citenamefont {Tseng}, \citenamefont
  {Ralph},\ and\ \citenamefont {Buhrman}}]{Liu.2012}%
  \BibitemOpen
  \bibfield  {author} {\bibinfo {author} {\bibfnamefont {L.}~\bibnamefont
  {Liu}}, \bibinfo {author} {\bibfnamefont {C.-F.}\ \bibnamefont {Pai}},
  \bibinfo {author} {\bibfnamefont {Y.}~\bibnamefont {Li}}, \bibinfo {author}
  {\bibfnamefont {H.~W.}\ \bibnamefont {Tseng}}, \bibinfo {author}
  {\bibfnamefont {D.~C.}\ \bibnamefont {Ralph}}, \ and\ \bibinfo {author}
  {\bibfnamefont {R.~A.}\ \bibnamefont {Buhrman}},\ }\href {\doibase
  10.1126/science.1218197} {\bibfield  {journal} {\bibinfo  {journal}
  {Science}\ }\textbf {\bibinfo {volume} {336}},\ \bibinfo {pages} {555}
  (\bibinfo {year} {2012})}\BibitemShut {NoStop}%
\bibitem [{\citenamefont {Wang}\ \emph {et~al.}(2018)\citenamefont {Wang},
  \citenamefont {Cai}, \citenamefont {Zhu}, \citenamefont {Wang}, \citenamefont
  {Kan}, \citenamefont {Zhao}, \citenamefont {Cao}, \citenamefont {Wang},
  \citenamefont {Zhang}, \citenamefont {Zhang}, \citenamefont {Park},
  \citenamefont {Wang}, \citenamefont {Fert},\ and\ \citenamefont
  {Zhao}}]{Wang.2018}%
  \BibitemOpen
  \bibfield  {author} {\bibinfo {author} {\bibfnamefont {M.}~\bibnamefont
  {Wang}}, \bibinfo {author} {\bibfnamefont {W.}~\bibnamefont {Cai}}, \bibinfo
  {author} {\bibfnamefont {D.}~\bibnamefont {Zhu}}, \bibinfo {author}
  {\bibfnamefont {Z.}~\bibnamefont {Wang}}, \bibinfo {author} {\bibfnamefont
  {J.}~\bibnamefont {Kan}}, \bibinfo {author} {\bibfnamefont {Z.}~\bibnamefont
  {Zhao}}, \bibinfo {author} {\bibfnamefont {K.}~\bibnamefont {Cao}}, \bibinfo
  {author} {\bibfnamefont {Z.}~\bibnamefont {Wang}}, \bibinfo {author}
  {\bibfnamefont {Y.}~\bibnamefont {Zhang}}, \bibinfo {author} {\bibfnamefont
  {T.}~\bibnamefont {Zhang}}, \bibinfo {author} {\bibfnamefont
  {C.}~\bibnamefont {Park}}, \bibinfo {author} {\bibfnamefont {J.-P.}\
  \bibnamefont {Wang}}, \bibinfo {author} {\bibfnamefont {A.}~\bibnamefont
  {Fert}}, \ and\ \bibinfo {author} {\bibfnamefont {W.}~\bibnamefont {Zhao}},\
  }\href {\doibase 10.1038/s41928-018-0160-7} {\bibfield  {journal} {\bibinfo
  {journal} {Nature Electronics}\ }\textbf {\bibinfo {volume} {1}},\ \bibinfo
  {pages} {582} (\bibinfo {year} {2018})}\BibitemShut {NoStop}%
\bibitem [{\citenamefont {Cubukcu}\ \emph {et~al.}(2018)\citenamefont
  {Cubukcu}, \citenamefont {Boulle}, \citenamefont {Mikuszeit}, \citenamefont
  {Hamelin}, \citenamefont {Brächer}, \citenamefont {Lamard}, \citenamefont
  {Cyrille}, \citenamefont {Buda-Prejbeanu}, \citenamefont {Garello},
  \citenamefont {Miron}, \citenamefont {Klein}, \citenamefont {Loubens},
  \citenamefont {Naletov}, \citenamefont {Langer}, \citenamefont {Ocker},
  \citenamefont {Gambardella},\ and\ \citenamefont {Gaudin}}]{Cubukcu.2018}%
  \BibitemOpen
  \bibfield  {author} {\bibinfo {author} {\bibfnamefont {M.}~\bibnamefont
  {Cubukcu}}, \bibinfo {author} {\bibfnamefont {O.}~\bibnamefont {Boulle}},
  \bibinfo {author} {\bibfnamefont {N.}~\bibnamefont {Mikuszeit}}, \bibinfo
  {author} {\bibfnamefont {C.}~\bibnamefont {Hamelin}}, \bibinfo {author}
  {\bibfnamefont {T.}~\bibnamefont {Brächer}}, \bibinfo {author}
  {\bibfnamefont {N.}~\bibnamefont {Lamard}}, \bibinfo {author} {\bibfnamefont
  {M.~C.}\ \bibnamefont {Cyrille}}, \bibinfo {author} {\bibfnamefont
  {L.}~\bibnamefont {Buda-Prejbeanu}}, \bibinfo {author} {\bibfnamefont
  {K.}~\bibnamefont {Garello}}, \bibinfo {author} {\bibfnamefont {I.~M.}\
  \bibnamefont {Miron}}, \bibinfo {author} {\bibfnamefont {O.}~\bibnamefont
  {Klein}}, \bibinfo {author} {\bibfnamefont {G.~d.}\ \bibnamefont {Loubens}},
  \bibinfo {author} {\bibfnamefont {V.~V.}\ \bibnamefont {Naletov}}, \bibinfo
  {author} {\bibfnamefont {J.}~\bibnamefont {Langer}}, \bibinfo {author}
  {\bibfnamefont {B.}~\bibnamefont {Ocker}}, \bibinfo {author} {\bibfnamefont
  {P.}~\bibnamefont {Gambardella}}, \ and\ \bibinfo {author} {\bibfnamefont
  {G.}~\bibnamefont {Gaudin}},\ }\href {\doibase 10.1109/TMAG.2017.2772185}
  {\bibfield  {journal} {\bibinfo  {journal} {IEEE Transactions on Magnetics}\
  }\textbf {\bibinfo {volume} {54}},\ \bibinfo {pages} {9300204} (\bibinfo
  {year} {2018})}\BibitemShut {NoStop}%
\bibitem [{\citenamefont {Buttner}\ \emph {et~al.}(2017)\citenamefont
  {Buttner}, \citenamefont {Lemesh}, \citenamefont {Schneider}, \citenamefont
  {Pfau}, \citenamefont {Gunther}, \citenamefont {Hessing}, \citenamefont
  {Geilhufe}, \citenamefont {Caretta}, \citenamefont {Engel}, \citenamefont
  {Kruger}, \citenamefont {Viefhaus}, \citenamefont {Eisebitt},\ and\
  \citenamefont {Beach}}]{Buttner.2017}%
  \BibitemOpen
  \bibfield  {author} {\bibinfo {author} {\bibfnamefont {F.}~\bibnamefont
  {Buttner}}, \bibinfo {author} {\bibfnamefont {I.}~\bibnamefont {Lemesh}},
  \bibinfo {author} {\bibfnamefont {M.}~\bibnamefont {Schneider}}, \bibinfo
  {author} {\bibfnamefont {B.}~\bibnamefont {Pfau}}, \bibinfo {author}
  {\bibfnamefont {C.~M.}\ \bibnamefont {Gunther}}, \bibinfo {author}
  {\bibfnamefont {P.}~\bibnamefont {Hessing}}, \bibinfo {author} {\bibfnamefont
  {J.}~\bibnamefont {Geilhufe}}, \bibinfo {author} {\bibfnamefont
  {L.}~\bibnamefont {Caretta}}, \bibinfo {author} {\bibfnamefont
  {D.}~\bibnamefont {Engel}}, \bibinfo {author} {\bibfnamefont
  {B.}~\bibnamefont {Kruger}}, \bibinfo {author} {\bibfnamefont
  {J.}~\bibnamefont {Viefhaus}}, \bibinfo {author} {\bibfnamefont
  {S.}~\bibnamefont {Eisebitt}}, \ and\ \bibinfo {author} {\bibfnamefont
  {G.~S.~D.}\ \bibnamefont {Beach}},\ }\href {\doibase 10.1038/nnano.2017.178}
  {\bibfield  {journal} {\bibinfo  {journal} {Nature Nanotechnology}\ }\textbf
  {\bibinfo {volume} {12}},\ \bibinfo {pages} {1040} (\bibinfo {year}
  {2017})}\BibitemShut {NoStop}%
\bibitem [{\citenamefont {Woo}\ \emph {et~al.}(2018)\citenamefont {Woo},
  \citenamefont {Song}, \citenamefont {Zhang}, \citenamefont {Ezawa},
  \citenamefont {Zhou}, \citenamefont {Liu}, \citenamefont {Weigand},
  \citenamefont {Finizio}, \citenamefont {Raabe}, \citenamefont {Park},
  \citenamefont {Lee}, \citenamefont {Choi}, \citenamefont {Min}, \citenamefont
  {Koo},\ and\ \citenamefont {Chang}}]{Woo.2018}%
  \BibitemOpen
  \bibfield  {author} {\bibinfo {author} {\bibfnamefont {S.}~\bibnamefont
  {Woo}}, \bibinfo {author} {\bibfnamefont {K.~M.}\ \bibnamefont {Song}},
  \bibinfo {author} {\bibfnamefont {X.}~\bibnamefont {Zhang}}, \bibinfo
  {author} {\bibfnamefont {M.}~\bibnamefont {Ezawa}}, \bibinfo {author}
  {\bibfnamefont {Y.}~\bibnamefont {Zhou}}, \bibinfo {author} {\bibfnamefont
  {X.}~\bibnamefont {Liu}}, \bibinfo {author} {\bibfnamefont {M.}~\bibnamefont
  {Weigand}}, \bibinfo {author} {\bibfnamefont {S.}~\bibnamefont {Finizio}},
  \bibinfo {author} {\bibfnamefont {J.}~\bibnamefont {Raabe}}, \bibinfo
  {author} {\bibfnamefont {M.-C.}\ \bibnamefont {Park}}, \bibinfo {author}
  {\bibfnamefont {K.-Y.}\ \bibnamefont {Lee}}, \bibinfo {author} {\bibfnamefont
  {J.~W.}\ \bibnamefont {Choi}}, \bibinfo {author} {\bibfnamefont {B.-C.}\
  \bibnamefont {Min}}, \bibinfo {author} {\bibfnamefont {H.~C.}\ \bibnamefont
  {Koo}}, \ and\ \bibinfo {author} {\bibfnamefont {J.}~\bibnamefont {Chang}},\
  }\href {\doibase 10.1038/s41928-018-0070-8} {\bibfield  {journal} {\bibinfo
  {journal} {Nature Electronics}\ }\textbf {\bibinfo {volume} {1}},\ \bibinfo
  {pages} {288} (\bibinfo {year} {2018})}\BibitemShut {NoStop}%
\bibitem [{\citenamefont {Finizio}\ \emph {et~al.}(2019)\citenamefont
  {Finizio}, \citenamefont {Zeissler}, \citenamefont {Wintz}, \citenamefont
  {Mayr}, \citenamefont {Wessels}, \citenamefont {Huxtable}, \citenamefont
  {Burnell}, \citenamefont {Marrows},\ and\ \citenamefont
  {Raabe}}]{Finizio.2019}%
  \BibitemOpen
  \bibfield  {author} {\bibinfo {author} {\bibfnamefont {S.}~\bibnamefont
  {Finizio}}, \bibinfo {author} {\bibfnamefont {K.}~\bibnamefont {Zeissler}},
  \bibinfo {author} {\bibfnamefont {S.}~\bibnamefont {Wintz}}, \bibinfo
  {author} {\bibfnamefont {S.}~\bibnamefont {Mayr}}, \bibinfo {author}
  {\bibfnamefont {T.}~\bibnamefont {Wessels}}, \bibinfo {author} {\bibfnamefont
  {A.~J.}\ \bibnamefont {Huxtable}}, \bibinfo {author} {\bibfnamefont
  {G.}~\bibnamefont {Burnell}}, \bibinfo {author} {\bibfnamefont {C.~H.}\
  \bibnamefont {Marrows}}, \ and\ \bibinfo {author} {\bibfnamefont
  {J.}~\bibnamefont {Raabe}},\ }\href {\doibase 10.1021/acs.nanolett.9b02840}
  {\bibfield  {journal} {\bibinfo  {journal} {Nano Letters}\ }\textbf {\bibinfo
  {volume} {19}},\ \bibinfo {pages} {7246} (\bibinfo {year}
  {2019})}\BibitemShut {NoStop}%
\bibitem [{\citenamefont {Bhattacharya}\ \emph {et~al.}(2020)\citenamefont
  {Bhattacharya}, \citenamefont {Razavi}, \citenamefont {Wu}, \citenamefont
  {Dai}, \citenamefont {Wang},\ and\ \citenamefont
  {Atulasimha}}]{Bhattacharya.2020}%
  \BibitemOpen
  \bibfield  {author} {\bibinfo {author} {\bibfnamefont {D.}~\bibnamefont
  {Bhattacharya}}, \bibinfo {author} {\bibfnamefont {S.~A.}\ \bibnamefont
  {Razavi}}, \bibinfo {author} {\bibfnamefont {H.}~\bibnamefont {Wu}}, \bibinfo
  {author} {\bibfnamefont {B.}~\bibnamefont {Dai}}, \bibinfo {author}
  {\bibfnamefont {K.~L.}\ \bibnamefont {Wang}}, \ and\ \bibinfo {author}
  {\bibfnamefont {J.}~\bibnamefont {Atulasimha}},\ }\href {\doibase
  10.1038/s41928-020-0432-x} {\bibfield  {journal} {\bibinfo  {journal} {Nature
  Electronics}\ }\textbf {\bibinfo {volume} {3}},\ \bibinfo {pages} {539}
  (\bibinfo {year} {2020})}\BibitemShut {NoStop}%
\bibitem [{\citenamefont {Wang}\ \emph {et~al.}(2012)\citenamefont {Wang},
  \citenamefont {Li}, \citenamefont {Hageman},\ and\ \citenamefont
  {Chien}}]{Wang.2012}%
  \BibitemOpen
  \bibfield  {author} {\bibinfo {author} {\bibfnamefont {W.-G.}\ \bibnamefont
  {Wang}}, \bibinfo {author} {\bibfnamefont {M.}~\bibnamefont {Li}}, \bibinfo
  {author} {\bibfnamefont {S.}~\bibnamefont {Hageman}}, \ and\ \bibinfo
  {author} {\bibfnamefont {C.~L.}\ \bibnamefont {Chien}},\ }\href {\doibase
  10.1038/nmat3171} {\bibfield  {journal} {\bibinfo  {journal} {Nature
  Materials}\ }\textbf {\bibinfo {volume} {11}},\ \bibinfo {pages} {64}
  (\bibinfo {year} {2012})}\BibitemShut {NoStop}%
\bibitem [{\citenamefont {Niranjan}\ \emph {et~al.}(2010)\citenamefont
  {Niranjan}, \citenamefont {Duan}, \citenamefont {Jaswal},\ and\ \citenamefont
  {Tsymbal}}]{Niranjan.2010}%
  \BibitemOpen
  \bibfield  {author} {\bibinfo {author} {\bibfnamefont {M.~K.}\ \bibnamefont
  {Niranjan}}, \bibinfo {author} {\bibfnamefont {C.-G.}\ \bibnamefont {Duan}},
  \bibinfo {author} {\bibfnamefont {S.~S.}\ \bibnamefont {Jaswal}}, \ and\
  \bibinfo {author} {\bibfnamefont {E.~Y.}\ \bibnamefont {Tsymbal}},\ }\href
  {\doibase 10.1063/1.3443658} {\bibfield  {journal} {\bibinfo  {journal}
  {Applied Physics Letters}\ }\textbf {\bibinfo {volume} {96}},\ \bibinfo
  {pages} {222504} (\bibinfo {year} {2010})}\BibitemShut {NoStop}%
\bibitem [{\citenamefont {Li}\ \emph {et~al.}(2017)\citenamefont {Li},
  \citenamefont {Fitzell}, \citenamefont {Wu}, \citenamefont {Karaba},
  \citenamefont {Buditama}, \citenamefont {Yu}, \citenamefont {Wong},
  \citenamefont {Altieri}, \citenamefont {Grezes}, \citenamefont {Kioussis},
  \citenamefont {Tolbert}, \citenamefont {Zhang}, \citenamefont {Chang},
  \citenamefont {Khalili~Amiri},\ and\ \citenamefont {Wang}}]{Li.2017}%
  \BibitemOpen
  \bibfield  {author} {\bibinfo {author} {\bibfnamefont {X.}~\bibnamefont
  {Li}}, \bibinfo {author} {\bibfnamefont {K.}~\bibnamefont {Fitzell}},
  \bibinfo {author} {\bibfnamefont {D.}~\bibnamefont {Wu}}, \bibinfo {author}
  {\bibfnamefont {C.~T.}\ \bibnamefont {Karaba}}, \bibinfo {author}
  {\bibfnamefont {A.}~\bibnamefont {Buditama}}, \bibinfo {author}
  {\bibfnamefont {G.}~\bibnamefont {Yu}}, \bibinfo {author} {\bibfnamefont
  {K.~L.}\ \bibnamefont {Wong}}, \bibinfo {author} {\bibfnamefont
  {N.}~\bibnamefont {Altieri}}, \bibinfo {author} {\bibfnamefont
  {C.}~\bibnamefont {Grezes}}, \bibinfo {author} {\bibfnamefont
  {N.}~\bibnamefont {Kioussis}}, \bibinfo {author} {\bibfnamefont
  {S.}~\bibnamefont {Tolbert}}, \bibinfo {author} {\bibfnamefont
  {Z.}~\bibnamefont {Zhang}}, \bibinfo {author} {\bibfnamefont {J.~P.}\
  \bibnamefont {Chang}}, \bibinfo {author} {\bibfnamefont {P.}~\bibnamefont
  {Khalili~Amiri}}, \ and\ \bibinfo {author} {\bibfnamefont {K.~L.}\
  \bibnamefont {Wang}},\ }\href {\doibase 10.1063/1.4975160} {\bibfield
  {journal} {\bibinfo  {journal} {Applied Physics Letters}\ }\textbf {\bibinfo
  {volume} {110}},\ \bibinfo {pages} {052401} (\bibinfo {year}
  {2017})}\BibitemShut {NoStop}%
\bibitem [{\citenamefont {Hsu}\ \emph {et~al.}(2017)\citenamefont {Hsu},
  \citenamefont {Kubetzka}, \citenamefont {Finco}, \citenamefont {Romming},
  \citenamefont {von Bergmann},\ and\ \citenamefont {Wiesendanger}}]{Hsu.2017}%
  \BibitemOpen
  \bibfield  {author} {\bibinfo {author} {\bibfnamefont {P.~J.}\ \bibnamefont
  {Hsu}}, \bibinfo {author} {\bibfnamefont {A.}~\bibnamefont {Kubetzka}},
  \bibinfo {author} {\bibfnamefont {A.}~\bibnamefont {Finco}}, \bibinfo
  {author} {\bibfnamefont {N.}~\bibnamefont {Romming}}, \bibinfo {author}
  {\bibfnamefont {K.}~\bibnamefont {von Bergmann}}, \ and\ \bibinfo {author}
  {\bibfnamefont {R.}~\bibnamefont {Wiesendanger}},\ }\href {\doibase
  10.1038/nnano.2016.234} {\bibfield  {journal} {\bibinfo  {journal} {Nature
  Nanotechnology}\ }\textbf {\bibinfo {volume} {12}},\ \bibinfo {pages} {123}
  (\bibinfo {year} {2017})}\BibitemShut {NoStop}%
\bibitem [{\citenamefont {Schott}\ \emph {et~al.}(2017)\citenamefont {Schott},
  \citenamefont {Bernand-Mantel}, \citenamefont {Ranno}, \citenamefont
  {Pizzini}, \citenamefont {Vogel}, \citenamefont {Béa}, \citenamefont
  {Baraduc}, \citenamefont {Auffret}, \citenamefont {Gaudin},\ and\
  \citenamefont {Givord}}]{Schott.2017}%
  \BibitemOpen
  \bibfield  {author} {\bibinfo {author} {\bibfnamefont {M.}~\bibnamefont
  {Schott}}, \bibinfo {author} {\bibfnamefont {A.}~\bibnamefont
  {Bernand-Mantel}}, \bibinfo {author} {\bibfnamefont {L.}~\bibnamefont
  {Ranno}}, \bibinfo {author} {\bibfnamefont {S.}~\bibnamefont {Pizzini}},
  \bibinfo {author} {\bibfnamefont {J.}~\bibnamefont {Vogel}}, \bibinfo
  {author} {\bibfnamefont {H.}~\bibnamefont {Béa}}, \bibinfo {author}
  {\bibfnamefont {C.}~\bibnamefont {Baraduc}}, \bibinfo {author} {\bibfnamefont
  {S.}~\bibnamefont {Auffret}}, \bibinfo {author} {\bibfnamefont
  {G.}~\bibnamefont {Gaudin}}, \ and\ \bibinfo {author} {\bibfnamefont
  {D.}~\bibnamefont {Givord}},\ }\href {\doibase 10.1021/acs.nanolett.7b00328}
  {\bibfield  {journal} {\bibinfo  {journal} {Nano Letters}\ }\textbf {\bibinfo
  {volume} {17}},\ \bibinfo {pages} {3006} (\bibinfo {year}
  {2017})}\BibitemShut {NoStop}%
\bibitem [{\citenamefont {Zhang}\ \emph {et~al.}(2022)\citenamefont {Zhang},
  \citenamefont {Bapna}, \citenamefont {Jiang}, \citenamefont {Sousa},
  \citenamefont {Liao}, \citenamefont {Zhao}, \citenamefont {Lv}, \citenamefont
  {Sahu}, \citenamefont {Lyu}, \citenamefont {Naeemi}, \citenamefont {Low},
  \citenamefont {Majetich},\ and\ \citenamefont {Wang}}]{Zhang.2022}%
  \BibitemOpen
  \bibfield  {author} {\bibinfo {author} {\bibfnamefont {D.}~\bibnamefont
  {Zhang}}, \bibinfo {author} {\bibfnamefont {M.}~\bibnamefont {Bapna}},
  \bibinfo {author} {\bibfnamefont {W.}~\bibnamefont {Jiang}}, \bibinfo
  {author} {\bibfnamefont {D.}~\bibnamefont {Sousa}}, \bibinfo {author}
  {\bibfnamefont {Y.-C.}\ \bibnamefont {Liao}}, \bibinfo {author}
  {\bibfnamefont {Z.}~\bibnamefont {Zhao}}, \bibinfo {author} {\bibfnamefont
  {Y.}~\bibnamefont {Lv}}, \bibinfo {author} {\bibfnamefont {P.}~\bibnamefont
  {Sahu}}, \bibinfo {author} {\bibfnamefont {D.}~\bibnamefont {Lyu}}, \bibinfo
  {author} {\bibfnamefont {A.}~\bibnamefont {Naeemi}}, \bibinfo {author}
  {\bibfnamefont {T.}~\bibnamefont {Low}}, \bibinfo {author} {\bibfnamefont
  {S.~A.}\ \bibnamefont {Majetich}}, \ and\ \bibinfo {author} {\bibfnamefont
  {J.-P.}\ \bibnamefont {Wang}},\ }\href {\doibase
  10.1021/acs.nanolett.1c03395} {\bibfield  {journal} {\bibinfo  {journal}
  {Nano Letters}\ }\textbf {\bibinfo {volume} {22}},\ \bibinfo {pages} {622}
  (\bibinfo {year} {2022})}\BibitemShut {NoStop}%
\bibitem [{\citenamefont {Chen}\ \emph {et~al.}(2023)\citenamefont {Chen},
  \citenamefont {Tai}, \citenamefont {Tan}, \citenamefont {Tan}, \citenamefont
  {Lim}, \citenamefont {Ho},\ and\ \citenamefont
  {Soumyanarayanan}}]{Chen.2023}%
  \BibitemOpen
  \bibfield  {author} {\bibinfo {author} {\bibfnamefont {X.}~\bibnamefont
  {Chen}}, \bibinfo {author} {\bibfnamefont {T.}~\bibnamefont {Tai}}, \bibinfo
  {author} {\bibfnamefont {H.}~\bibnamefont {Tan}}, \bibinfo {author}
  {\bibfnamefont {H.}~\bibnamefont {Tan}}, \bibinfo {author} {\bibfnamefont
  {R.}~\bibnamefont {Lim}}, \bibinfo {author} {\bibfnamefont {P.}~\bibnamefont
  {Ho}}, \ and\ \bibinfo {author} {\bibfnamefont {A.}~\bibnamefont
  {Soumyanarayanan}},\ }\href@noop {} {\bibfield  {journal} {\bibinfo
  {journal} {ArXiv e-prints}\ } (\bibinfo {year} {2023})},\ \Eprint
  {http://arxiv.org/abs/2301.07327} {arXiv:2301.07327 [cond-mat.mtrl-sci]}
  \BibitemShut {NoStop}%
\bibitem [{\citenamefont {Vansteenkiste}\ \emph {et~al.}(2014)\citenamefont
  {Vansteenkiste}, \citenamefont {Leliaert}, \citenamefont {Dvornik},
  \citenamefont {Helsen}, \citenamefont {Garcia-Sanchez},\ and\ \citenamefont
  {Van~Waeyenberge}}]{Vansteenkiste.2014}%
  \BibitemOpen
  \bibfield  {author} {\bibinfo {author} {\bibfnamefont {A.}~\bibnamefont
  {Vansteenkiste}}, \bibinfo {author} {\bibfnamefont {J.}~\bibnamefont
  {Leliaert}}, \bibinfo {author} {\bibfnamefont {M.}~\bibnamefont {Dvornik}},
  \bibinfo {author} {\bibfnamefont {M.}~\bibnamefont {Helsen}}, \bibinfo
  {author} {\bibfnamefont {F.}~\bibnamefont {Garcia-Sanchez}}, \ and\ \bibinfo
  {author} {\bibfnamefont {B.}~\bibnamefont {Van~Waeyenberge}},\ }\href
  {\doibase 10.1063/1.4899186} {\bibfield  {journal} {\bibinfo  {journal} {AIP
  Advances}\ }\textbf {\bibinfo {volume} {4}},\ \bibinfo {pages} {107133}
  (\bibinfo {year} {2014})}\BibitemShut {NoStop}%
\end{thebibliography}%

\vspace{0.1ex}
\begin{center}
\rule[0.5ex]{0.5\columnwidth}{0.2pt}
\par\end{center}
\vspace{0.1ex}

\textsf{\textbf{\small{}Acknowledgments.}}{\small{}
We acknowledge helpful discussions with Franck Ernult and Bingjin Chen, and experimental inputs from Jinjun Qiu, Prerna Chauhan, Zeyu Ma, and Yangwenxuan Niu. 
This work was supported by the SpOT-LITE programme (Grant No. A18A6b0057), funded by Singapore's RIE2020 initiatives.}

\end{document}